\documentclass[useAMS,usenatbib]{mn2e}
\usepackage[dvips]{graphicx}
\usepackage{epstopdf}
\usepackage{amssymb}

\newcommand{\etal}{~et~al.}

\def\aj{AJ}%
\def\apj{ApJ}%
\def\apjl{ApJ}%
\def\apjs{ApJS}%
\def\aap{A\&A}%
\def\mnras{MNRAS}%
\title[NIR spectroscopy of post-starburst galaxies]{Near Infrared
  spectroscopy of post-starburst galaxies: a limited
  impact of TP-AGB stars on galaxy SEDs\thanks{Based on observations made with ESO
    telescopes at the La Silla Paranal Observatory under programme ID
    086.B-0733.}}

\author[S. Zibetti\etal]{Stefano Zibetti$^{1,2}$\thanks{E-mail:
    zibetti@arcetri.astro.it}, Anna Gallazzi$^{2}$, St\'ephane
  Charlot$^{3}$,
  Daniele Pierini\thanks{Free-lance scientist}, Anna Pasquali$^{4}$\\
  $^{1}$INAF-Osservatorio Astrofisico di Arcetri - Largo Enrico Fermi, 5 - I-50125 Firenze, Italy\\
  $^{2}$Dark Cosmology Centre, Niels Bohr Institute - University of
  Copenhagen Juliane Maries Vej 30, DK-2100 Copenhagen, Denmark\\
  $^3$Institut d’Astrophysique de Paris, CNRS, Universit\'e Pierre \& Marie Curie, 98 bis Boulevard Arago, 75014 Paris, France \\
  $^4$Astronomisches Rechen-Institut, Zentrum f\"ur Astronomie der
  Universit\"at Heidelberg, M\"onchhofstrasse 12 - 14, 69120 Heidelberg,
  Germany }
\begin{document}
\citestyle{mn2e}
\bibliographystyle{mn2e}

\date{Accepted 2012 October 2.  Received 2012 September 4; in original form 2012 May 20.}

\pagerange{\pageref{firstpage}--\pageref{lastpage}} \pubyear{2012}

\maketitle

  \label{firstpage}

\begin{abstract}
  We present VLT-ISAAC near infrared (NIR) spectro-photometric
  observations of 16 post-starburst galaxies aimed at constraining the
  debated influence of TP-AGB stars on the spectral energy
  distribution (SED) of galaxies with stellar ages between 0.5 and 2
  Gyr, hence critical for high-redshift studies. Post-starburst
  galaxies are characterised by negligible on-going star formation and
  a SED dominated by the stellar population formed in a recent ($<2$
  Gyr) burst. By selecting post-starburst galaxies with mean
  luminosity-weighted ages between 0.5 and 1.5 Gyr and a broad range
  of metallicities (based on SDSS optical spectroscopy), we explore
  the parameter space over which the relative energy output of TP-AGB
  stars peaks. A key feature of the present study is that we target
  galaxies at $z\approx0.2$, so that two main spectral features of
  TP-AGB stars (C-molecule band-head drops at 1.41 and 1.77~$\mu$m,
  blended with strong telluric absorption features, hence hardly
  observable from the ground, for targets at $z\approx0$) move inside
  the H and K atmospheric windows and can be constrained for the first
  time to high accuracy. Our observations provide key constraints to
  stellar population synthesis models. Our main results are:
  \textit{i)} the NIR regions around 1.41 and 1.77~$\mu$m (rest-frame)
  are featureless for all galaxies in our sample over the whole range
  of relevant ages and metallicities at variance with the Maraston
  (2005) ``TP-AGB heavy'' models, which exhibit marked drops there;
  \textit{ii)} no flux boosting is observed in the NIR. The
  optical-NIR SEDs of most of our post-starburst galaxies can be
  consistently reproduced with Bruzual \& Charlot (2003) models, using
  either simple stellar populations (SSP) of corresponding
  light-weighted ages and metallicities, or a more realistic burst
  plus an underlying old population containing up to approximately
  60\% of the total stellar mass. In contrast, all combinations of
  this kind based on the Maraston (2005) models are unable to
  simultaneously reproduce the smoothness of the NIR spectra and the
  relatively blue optical-NIR colours in the observations. The data
  collected in this study appear to disfavour "TP-AGB heavy" models
  with respect to "TP-AGB light" ones.
\end{abstract}


\section{Introduction}\label{sec:intro}

The modelling of the spectral energy distribution (SED) of galaxies
and its interpretation in terms of stellar population parameters
(e.g. age, star formation history, metallicity) via stellar population
synthesis (SPS) is a fundamental tool to understand the properties and
evolution of galaxies. Yet, we are far from a complete comprehension
and a reliable modelling of some stellar evolutionary phases which
strongly affect the energy output of stellar populations. Among them,
the so-called thermally pulsing - asymptotic giant branch (TP-AGB)
phase has been the focus of debate among modellers for several years
(e.g. \citealt{charlot_bruzual_91,maraston05,bruzual07,marigo+08}). This
late stage of the AGB phase for low and intermediate mass stars
($M\lesssim$ 5--7 M$_\odot$) culminates in stellar populations of ages
between 0.5 and 1.5 Gyr. TP-AGB stars emit mainly in the near infrared
(NIR) spectral range given the low temperatures of these
stars. Quantitative predictions of the lifetimes, luminosities and
spectral shapes and features of these stars as well as of their impact
on the integrated spectra of even ``simple'' stellar populations, are
still very uncertain and strongly model-dependent
\citep[e.g.][]{melbourne+12}. One further complication derives from
the fact that TP-AGB stars are likely to be embedded in envelopes of
dust produced by themselves
\citep[e.g.][]{groenewegen+09,chisari_kelson12}, which attenuates the
emerging optical and NIR radiation by a substantial fraction
\citep{meidt+12}.

\begin{figure*}
\includegraphics[width=\textwidth]{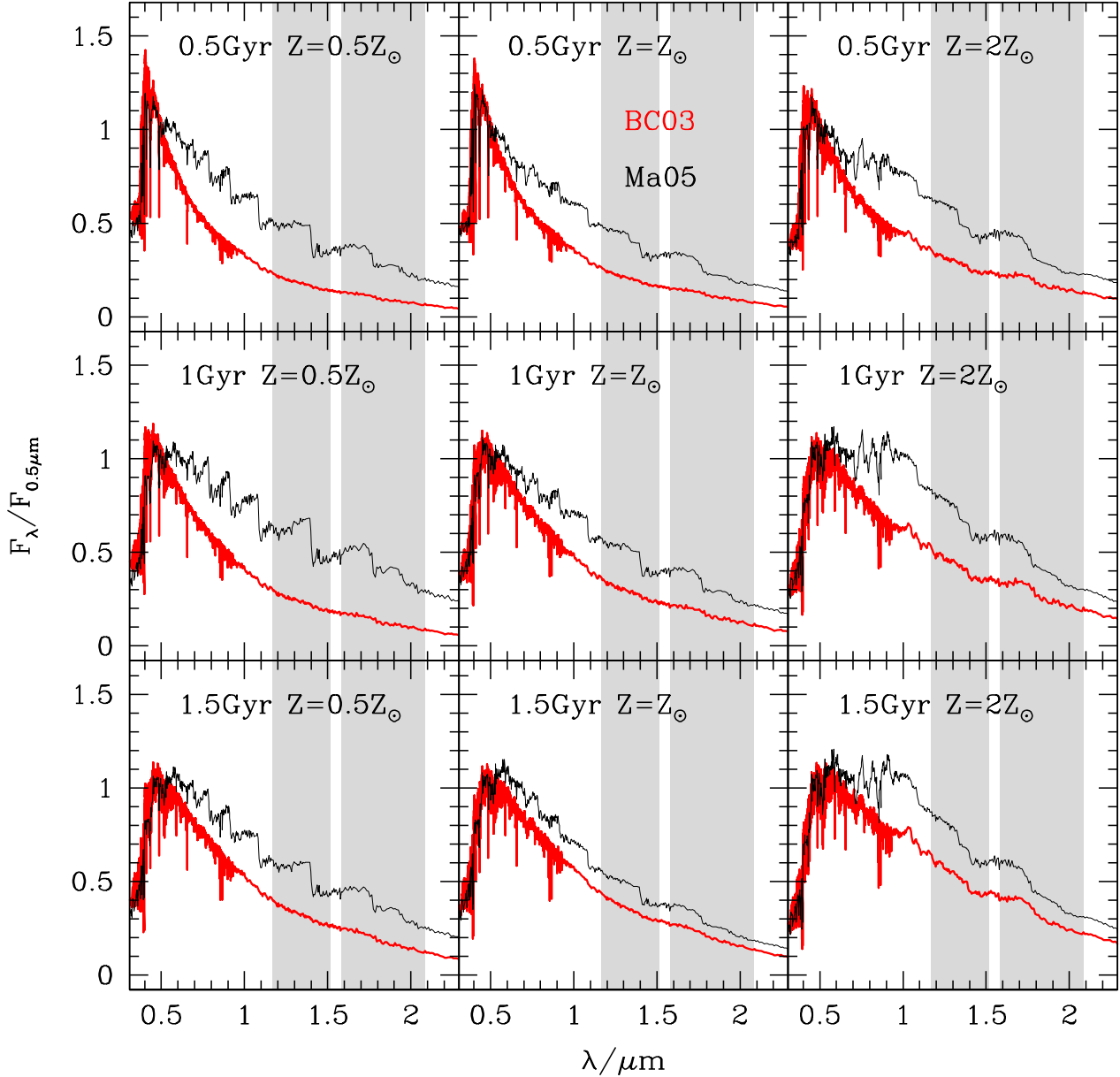}
\caption{Comparison between SSPs at the peak of the TP-AGB phase, as
  derived from Ma05 models (black lines) and BC03 (red lines). Three
  different ages, 0.5, 1 and 1.5 Gyr from top to bottom, and three
  different metallicities, 0.5, 1 and 2 $\mathrm{Z_\odot}$ (the
  highest metallicity for BC03 is actually replaced by 2.5
  $\mathrm{Z_\odot}$), from left to right, are considered. Shaded
  regions correspond to the rest-frame spectral ranges covered by the
  H and K band ISAAC spectroscopic setups for the $z\approx 0.2$
  galaxies presented in this work and include the strong features at
  1.41 and 1.77~$\mu$m.}\label{fig:Ma05vsBC03}
\end{figure*}

The seminal work of \citet[][Ma05 hereafter]{maraston05}, who adopted
the ``fuel consumption theorem'' approach and calibrated the flux
contribution of this phase against optical and NIR photometry of
Magellanic Clouds' globular clusters, dramatically pointed out how
large the impact of TP-AGB stars on galaxy SEDs may possibly be. The
Ma05 models make two clear and testable predictions in contrast to
``classical'' models \citep[such as][BC03 hereafter]{bc03}, which rely
on standard, less extreme prescriptions for the TP-AGB phase: in
particular, \textit{i)} the NIR flux of stellar populations of ages
between $\approx$0.5 and $\approx 1.5$ Gyr is enhanced by a factor up
to 3 (as a function of age and metallicity); in addition, \textit{ii)}
strong, sharp absorption features appear in the spectrum, especially
at NIR wavelengths (and most notably at 1.1, 1.41 and
1.77~$\mu$m). These features correspond to the band-heads of carbon
composite molecules and should depend on metallicity. They represent
the undebatable fingerprints of TP-AGB stars as predicted by Ma05
models. This can be clearly seen in Fig. \ref{fig:Ma05vsBC03}, where
we confront the optical-NIR spectra of simple stellar populations
(SSP) of three ages (0.5, 1 and 1.5 Gyr) at three different
metallicities (0.5, 1 and 2--2.5 $\mathrm{Z_\odot}$) as derived from
the Ma05 models (black lines) and the BC03 models (red lines)
respectively.

Observationally, detections of near-infrared CN absorption bands have
been reported in the integrated spectra of only a few Seyfert galaxies
\citep{riffel+07} and globular clusters dominated by the stochastic
presence of bright carbon stars \citep{lyubenova+12}. In fact, from a
theoretical point of view, carbon absorption features are not a
required signature of AGB-star populations. Sophisticated models of
AGB stellar evolution predict that convective dredge-up of carbon
produced by nuclear fusion during the envelope thermal pulses will
increase the abundance of carbon relative to oxygen near the surface,
which can eventually lead to the production of a carbon star
\citep[e.g.][]{marigo_girardi07}. While the efficiency of this process
is expected to increase as the metallicity decreases, the integrated
spectrum of most intermediate-age stellar populations could well be
dominated oxygen-rich AGB stars without strong carbon absorption
features.

The largely different NIR flux predictions for TP-AGB stars by
different models have strong implications in terms of SED
interpretation, leading to substantially lower stellar mass and age
estimates (by up to a factor 2 or more) when rest-frame NIR bands are
included in the analysis of galaxies with ages around 1 Gyr
\citep[e.g.][]{maraston05,maraston+06,kannappan_gawiser_07,conroy+09,ilbert+10}. The
popularity of NIR passbands as tracers of stellar mass, both at low
and high redshift, thanks to their (generally) low sensitivity to age,
metallicity and dust in terms of mass-to-light ratio (M/L)
\citep[e.g.][]{bell+03,ZCR09}, clearly calls for a proper assessment
of the models in this respect \citep[see
also][]{melbourne+12}. Verifying the reliability of SPS models in the
regime of ages around 1 Gyr is therefore of paramount relevance for
studies of galaxy properties and their evolution.

Since 2005, a number of contrasting results and claims have been
produced about the appropriateness of the Ma05
predictions. \cite{maraston+06} showed that the Ma05 models fit the
SEDs of high-z ($1.5\lesssim z \lesssim 2.5$) galaxies better than
BC03 models. However the same work shows that the performance of the
two models differ only marginally in terms of goodness of fit if one
allows for dust reddening, although the inferred stellar mass is
substantially different. A better agreement of Ma05 models with the
observed optical-NIR colours is also reported by \cite{macarthur+10},
who used combined optical spectroscopy and optical-NIR photometry to
constrain the star formation history of two
galaxies. \cite{lyubenova+12} also report detections of the 1.77~$\mu$m
feature in two Magellanic Clouds' globular clusters, in agreement with
Ma05 predictions. On the other hand, studies of Magellanic Clouds'
globular clusters found significant discrepancies between observations
and Ma05 predictions. \cite{lyubenova+10} found that the CO line
strength (in the NIR K band) is much weaker than predicted by Ma05 in
clusters of age $\approx$ 1 Gyr. \cite{conroy_gunn_10} studied the
colours of Magellanic Clouds' globular clusters as a function of their
age and showed that Ma05 models predict optical and NIR colours that
are too red. Moreover the predicted age dependence of these colours
does not agree with the data. As opposed, a substantial agreement with
models based on much less extreme TP-AGB prescriptions is found.

As already pointed out by \cite{lancon+99}, post-starburst galaxies
(PSB hereafter) are the best galaxies to look at in order to find the
fingerprints of (TP-)AGB stars on their SED. PSBs have a star
formation history dominated by a burst in the exact age range when
TP-AGB stars are expected to have the strongest impact, i.e. between
0.5 and 1.5 Gyr prior to the epoch of observation\footnote{This holds
  for several different star formation histories \citep[see figs. 11
  and 12 in][]{melbourne+12}.}. Their optical spectrum is
characterised by a well developed Balmer/D4000 break, the lack of
strong emission lines (which testify a substantial cessation of star
formation activity) and by the very strong Balmer absorption lines
typical of A stars (hence the alternative denomination of E+A or K+A
galaxies). \cite{conroy_gunn_10} compared the optical-NIR colours of
the sample of PSBs spectroscopically selected by \cite{quintero+04} to
the colours of different SPS models. They showed (see their Fig. 16)
that $i-z$, $g-r$ colours are not reproduced by Ma05 models: not only
would the blue portion of this colour plane be only explained by
unrealistically low metallicities, but also the age sequences in Ma05
models produce trends that are orthogonal to the observed age trends.
\cite{kriek+10} performed another photometric test, in which they
fitted the stacked (rest-frame) UV-optical-NIR SED of 62 PSBs at $0.7
\lesssim z \lesssim 2$, colour selected from the \textit{NEWFIRM}
medium band survey (NMBS) to have ages around 1 Gyr, according to both
BC03 and Ma05 models. Also in this case it is found that BC03 models
can reproduce the observed SEDs much better, in particular the
observed blue optical-NIR colours, which are not reproduced by Ma05.

In this paper we present the results of the experiment that we
designed to test for the two main predictions of Ma05 models, namely
the presence of sharp NIR spectral features and the boosted NIR flux,
on a sample of PSBs carefully selected from optical spectroscopy and
newly observed in NIR spectroscopy (H and K band) with ISAAC at the
ESO-VLT. The selection criteria and properties of the sample are fully
detailed in Section \ref{sec:sample_obs}. Here we would like to stress
the uniqueness of this work. \textit{i)} For the first time we cover
both the optical and the NIR range of PSBs with flux calibrated
spectra to look directly for the NIR spectral features at 1.41 and
1.77~$\mu$m predicted by the Ma05 models and resulting from the large
impact of TP-AGB C-rich stars. These features are unaccessible or
extremely difficult to measure from the ground as they roughly
coincide with the atmospheric gaps between J and H, and between H and
K bands, respectively, for nearby objects. Therefore we observed PSBs
in the redshift range 0.15--0.25 so that the C-features move well into
the observable windows and are minimally affected by uncertain
telluric corrections, as shown in
Fig. \ref{fig:Ma05vsBC03}. \textit{ii)} Contrary to previous works
based either on colour selection \citep{kriek+10} or on a purely
phenomenological spectroscopic selection \citep{conroy_gunn_10}, the
selection of our sample is based on both phenomenological criteria
\citep[high Balmer line strength and low emission lines,][]{goto_05}
and bayesian estimates of light weighted ages from optical spectral
indices as in \cite{gallazzi+05}. This allows us to focus our
observations and analysis on the \textit{precise} range of stellar
ages where TP-AGB stars are expected to contribute the most, for a
variety of stellar metallicities.

The paper is organised as follows. In Section \ref{sec:sample_obs} we
present the sample, the new ISAAC NIR spectroscopic observations and
the data reduction, including spectrophotometric calibrations and
integration with optical data. In Section \ref{sec:analysis} we
present the results and discuss possible spurious effects and
contaminations that might affect our conclusions. A comparison with
the analysis of \cite{kriek+10} is also presented. Finally, in Section
\ref{sec:summary} we summarise our findings and propose our
conclusions and future developments.

\section{Sample and observations}\label{sec:sample_obs}
\subsection{Sample selection and characterisation}
We draw our PSB targets from the spectroscopic sample of the Sloan
Digital Sky Survey \citep[SDSS,][]{SDSS,strauss_etal02} data release 7
\citep[DR7,][]{SDSS_DR7}. The initial pre-selection was based on the
line equivalent width (EW) criteria defined in \cite{goto_05}, namely:
EW(H$\delta$)$>5$~\AA, EW(H$\alpha$)$>-3$~\AA,
EW([O\,II])$>-2.5$~\AA.\footnote{We adopt the convention of positive
  EW for absorption and negative EW for emission. For the [O\,II]
  emission, the sum of the doublet is considered.}  In order to ensure
that the two strong NIR features corresponding to the band-heads of CN
(1.41~$\mu$m) and C$_2$ (1.77~$\mu$m) due to carbon rich AGB stars
\citep[cf.][]{lancon_mouhcine_02} are observable from the ground, we
require the galaxies to be at $0.15<z<0.25$, so that 1.41~$\mu$m and
1.77~$\mu$m (at rest) move into the H- and K-band windows,
respectively. This selection is based on the SDSS-DR7 casjobs
catalog\footnote{\texttt http://casjobs.sdss.org/}. We further
restrict the sample by applying the visibility cuts in order for
targets to be observable from Paranal during the winter semester:
DEC$<20$ deg and RA between 0 and 230 deg or RA$> 330$~deg. Finally we
require that the stellar light-weighted age of the galaxies is in the
range 0.5 to 1.5 Gyr, i.e. at the maximum of the predicted TP-AGB
luminosity contribution, based on the estimates computed as in
\cite{gallazzi+05} and available in the MPA-JHU
catalog\footnote{\texttt http://www.mpa-garching.mpg.de/SDSS/. Note
  that although the MPA-JHU catalog is limited to DR4 only, no loss of
  possible targets derived from this cut.}.  Given the relative rarity
of the post starburst phase and the restrictive age, coordinate and
redshift cuts (the latter particularly relevant as it is almost at the
extreme boundary of the SDSS main galaxy sample), the final sample is
composed of sixteen galaxies only.

\begin{figure}
\includegraphics[width=\columnwidth]{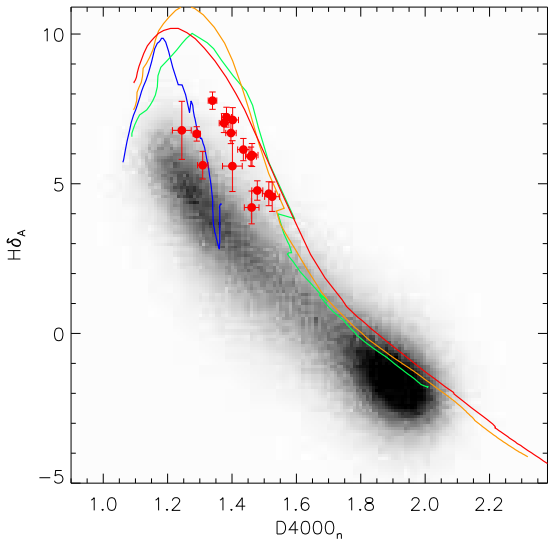}
\caption{Sample properties in the classical H$\delta$-D4000$_{\mathrm
    n}$ diagram, a proxy for burstiness vs. age of the stellar
  population. Red points with error bars are the sixteen galaxies
  selected for the present study; the underlying grey-scale image
  represents the density distribution of a complete sample of galaxies
  from SDSS-DR4. As one can see, our sample selects galaxies dominated
  by a burst in a range of young to intermediate ages. Note that the
  estimates of H$\delta_{\mathrm{A}}$ used in this plot are not
  exactly the same as the E.W. measurements used in the sample
  selection (therefore values $<5$~\AA~are not inconsistent with the
  selection criteria reported in the text).}\label{fig:sample_diagn}
\end{figure}

The properties of the sample are summarised in Table \ref{tab:sample}
and in Fig. \ref{fig:sample_diagn}. This figure illustrates the
distribution of our galaxies in the classical H$\delta_{\mathrm A}$ vs
$\mathrm{D4000_n}$ plane. In this plane the main sequence of galaxies
with smooth continuum star formation history is displayed by the grey
intensity levels, which represent the distribution of SDSS-DR4
galaxies of increasing ages from top-left to bottom right. Galaxies
lying above this sequence, such as our galaxies, represented by the
red dots with error bars, are characterised by relatively recent
bursts of star formation with ages between a few hundred Myr to a
couple of Gyr, which dominate the light of the galaxy. For comparison,
tracks of Simple Stellar Populations (SSP) with different
metallicities from BC03 are shown; the typical bell-shape of these SSP
tracks peaks in H$\delta_{\mathrm A}$ at ages of 0.3--0.5 Gyr
(depending on metallicity) and decreases to H$\delta_{\mathrm
  A}\approx$ 3--5 for ages of a few Gyr. By design, our sample
populates the intermediate range of burst ages in this diagram.

The statistical estimates of light-weighted stellar ages, metallicity
and mass computed from optical spectral absorption indices
($\mathrm{D4000_n}$, $\mathrm{H\delta_A+H\gamma_A}$,
$\mathrm{H\beta}$, $\mathrm{[MgFe]^\prime}$, $\mathrm{[Mg_2Fe]}$) with
the bayesian method of \cite{gallazzi+05} and used in the selection
are reported in Tab. \ref{tab:sample} (columns 6, 7 and 8), along with
the ID (1), coordinates (2, 3), redshift (4) and apparent $r$-band
petrosian magnitudes (5) of the galaxies from the SDSS. The sample
comprises intermediate mass galaxies, between $\approx 3$ and $\approx
7\times10^{10}\mathrm{M_\odot}$, with ages in the range $0.8-1.4$~Gyr
and metallicities between 0.2 and 2.3 solar. Most galaxies have indeed
$Z>0.5\mathrm{Z_\odot}$, as expected given their stellar mass. Despite
the paucity of galaxies at low age ($<1$~Gyr) \textit{and} low
metallicity ($<1\mathrm{Z_\odot}$), the sample provides a good
coverage of the relevant ages for the TP-AGB phase and of the
metallicities most typical for galaxies around the characteristic
luminosity $L^*$.
\begin{table*}
\begin{minipage}{\textwidth}
\caption{Sample properties.}\label{tab:sample}
\begin{tabular}{lrrrrrrr}
  \hline
  ID & RA & Dec  & z & $r_{\mathrm{petro}}$ & L-w Age & L-w $Z$ & M$^*$\\
  & (J2000.0)  & (J2000.0) & & mag & Gyr & $Z_\odot$ & $10^{10}\times$M$_\odot$      \\
  (1) & (2) & (3)  & (4) & (5) & (6) & (7) & (8)\\
  \hline
\smallskip
PSB\,J0151$-$0056 & 01:51:07.02 & $-$00:56:36.78 & 0.198  & 17.50 & $ 1.06_{-0.02}^{+0.02}$ & $ 2.31_{-0.35}^{+0.01}$ & $ 5.26_{-0.47}^{+0.15}$\\
\smallskip
PSB\,J0227$-$0015 & 02:27:43.21 & $-$00:15:23.08 & 0.219  & 16.92 & $ 1.04_{-0.08}^{+0.69}$ & $ 0.77_{-0.07}^{+0.01}$ & $ 7.06_{-0.41}^{+2.56}$\\
\smallskip
PSB\,J0328$+$0045 & 03:28:02.62 & $+$00:45:02.43 & 0.202  & 17.50 & $ 1.42_{-0.12}^{+0.02}$ & $ 1.68_{-0.01}^{+0.01}$ & $ 6.86_{-0.16}^{+0.17}$\\
\smallskip
PSB\,J0957$+$0249 & 09:57:29.91 & $+$02:49:42.02 & 0.216  & 17.37 & $ 0.89_{-0.01}^{+0.01}$ & $ 1.61_{-0.01}^{+0.01}$ & $ 4.96_{-0.11}^{+0.12}$\\
\smallskip
PSB\,J1006$+$1308 & 10:06:21.59 & $+$13:08:45.91 & 0.186  & 17.30 & $ 1.45_{-0.14}^{+0.29}$ & $ 0.37_{-0.05}^{+0.27}$ & $ 4.30_{-0.60}^{+0.74}$\\
\smallskip
PSB\,J1015$+$0103 & 10:15:19.69 & $+$01:03:41.68 & 0.216  & 17.22 & $ 0.89_{-0.01}^{+0.03}$ & $ 1.61_{-0.37}^{+0.01}$ & $ 5.31_{-0.13}^{+0.14}$\\
\smallskip
PSB\,J1039$+$0604 & 10:39:34.51 & $+$06:04:25.00 & 0.163  & 17.13 & $ 0.77_{-0.01}^{+0.01}$ & $ 2.21_{-0.01}^{+0.01}$ & $ 3.14_{-0.07}^{+0.07}$\\
\smallskip
PSB\,J1046$+$0714 & 10:46:19.33 & $+$07:14:52.34 & 0.173  & 17.59 & $ 1.44_{-0.17}^{+0.42}$ & $ 0.68_{-0.26}^{+0.44}$ & $ 3.42_{-0.23}^{+0.56}$\\
\smallskip
PSB\,J1119$+$1313 & 11:19:49.98 & $+$13:13:02.54 & 0.186  & 17.09 & $ 0.85_{-0.01}^{+0.01}$ & $ 2.15_{-0.01}^{+0.01}$ & $ 4.66_{-0.12}^{+0.12}$\\
\smallskip
PSB\,J1125$+$0049 & 11:25:55.17 & $+$00:49:58.82 & 0.181  & 17.43 & $ 1.03_{-0.02}^{+0.01}$ & $ 1.81_{-0.01}^{+0.01}$ & $ 5.66_{-0.14}^{+0.15}$\\
\smallskip
PSB\,J1133$+$0205 & 11:33:49.62 & $+$02:05:14.90 & 0.178  & 17.41 & $ 1.29_{-0.60}^{+0.54}$ & $ 0.62_{-0.29}^{+1.35}$ & $ 4.41_{-0.97}^{+0.77}$\\
\smallskip
PSB\,J1141$+$1014 & 11:41:15.54 & $+$10:14:25.68 & 0.215  & 17.63 & $ 1.11_{-0.26}^{+0.64}$ & $ 1.36_{-0.72}^{+0.79}$ & $ 4.43_{-0.46}^{+1.44}$\\
\smallskip
PSB\,J1150$-$0107 & 11:50:01.44 & $-$01:07:37.02 & 0.169  & 17.79 & $ 0.84_{-0.31}^{+0.70}$ & $ 0.21_{-0.13}^{+0.26}$ & $ 4.51_{-0.52}^{+1.91}$\\
\smallskip
PSB\,J1206$+$0918 & 12:06:30.81 & $+$09:18:55.09 & 0.179  & 17.58 & $ 1.27_{-0.37}^{+0.35}$ & $ 0.55_{-0.25}^{+0.72}$ & $ 2.67_{-0.45}^{+0.71}$\\
\smallskip
PSB\,J1230$+$1038 & 12:30:54.47 & $+$10:38:31.63 & 0.222  & 17.69 & $ 0.97_{-0.03}^{+2.34}$ & $ 1.96_{-0.02}^{+0.53}$ & $ 5.10_{-0.24}^{+5.62}$\\
\smallskip
PSB\,J1314$+$0430 & 13:14:14.26 & $+$04:30:47.65 & 0.159  & 17.14 & $ 1.27_{-0.09}^{+0.03}$ & $ 1.01_{-0.03}^{+0.01}$ & $ 3.79_{-0.20}^{+0.19}$\\
\hline
\end{tabular}
\end{minipage}
\end{table*}

We note that the estimates of the stellar population parameters of
\cite{gallazzi+05} rely on the comparison with synthetic spectral
libraries based on BC03 models, therefore they could be biased with
respect to estimates based on other synthesis models, most notably the
Ma05 ones. However, optical spectral indices short-wards of
5500~\AA~are generally well agreed upon among different modellers and,
most importantly, are only weakly affected by TP-AGB stars, which
mainly influence the NIR spectral region. Thus age and metallicity
estimates from optical absorption features are robust against
different TP-AGB prescriptions and the use of different models. This
is even more the case for spectra dominated by A stars, where Balmer
absorptions are very strong. Furthermore, in Sec. \ref{sec:analysis}
(\ref{sec:SSPcomp} and \ref{sec:CSPcomp} in particular) we thoroughly
explore if and how different (combinations of) ages and metallicities
can possibly explain the observed properties of our galaxies,
irrespective of their original selection.

Based on the SDSS optical images and, to minor extent, on our ISAAC
NIR images, we note that the majority of galaxies display smooth and
centrally concentrated light distributions, typical of early type
galaxies, with the only exception of PSB\,J0227$-$0015, which has an
extended and irregular ``halo''. However, other four galaxies
(PSB\,J1039$+$0604, PSB\,J1046$+$0714, PSB\,J1119$+$1313 and
PSB\,J1206$+$0918) have close and possibly interacting companions; in
many cases, faint structures (like shells or streamers) can be
tentatively seen. This suggests a relatively recent merger or
interaction as a possible origin of the ceased starburst in these
galaxies.

\subsection{Observations and data reduction}
The sixteen galaxies of the sample were observed in the programme ID
086.B-0733 (P.I.: S. Zibetti) at the UT3 Melipal of the ESO-VLT at the
Paranal Observatory. The ISAAC instrument \citep{ISAAC} has been used
in two different spectroscopic setups to obtain low resolution spectra
in the H and K bands. 1.5 arcsec slit and SWS1-LR grism were used in
both setups, with central wavelengths of 1.65 and 2.2~$\mu$m
respectively. This gives a spectral resolution of $\approx 350$
(50~\AA) and $\approx 300$ (70~\AA) in H and K
respectively. Observations were carried out in service mode with clear
(not necessarily photometric) sky and maximum seeing of 1.4 arcsec to
minimise slit losses and match the fibre size and seeing of the SDSS
optical spectra as well as possible. Standard ABBA on-slit nodding
with 70 arcsec offsets and 20 arcsec jittering were adopted to allow
accurate sky subtraction and bad pixels and cosmic ray
rejection. Eight exposures of 180 sec each were taken in H band and
six exposures of 150 sec each were taken in K, for total 24 minutes of
integration in H and 15 minutes in K per target. This eventually
yields typical signal-to-noise ratio (SNR) of $\approx 50$ and 25 per
resolution element in H and K respectively.  Spectro-photometric
standard stars were observed with identical setup immediately after or
before the science observations to provide a reliable estimate of the
spectral sensitivity function and compute the corrections for telluric
absorption. Standard data reduction procedures have been implemented
in IRAF\footnote{IRAF is distributed by the National Optical Astronomy
  Observatory, which is operated by the Association of Universities
  for Research in Astronomy (AURA) under cooperative agreement with
  the National Science Foundation.}. Wavelength calibration has been
performed based on the sky OH lines as measured on a median combined
image of the scientific frames. This approach turned out to be much
more stable and reliable than using the arc lamp calibrations taken in
day time. We have tried different approaches to optimally correct the
spectra for telluric absorption. As our best solution we adopted the
following strategy: the empirical high-resolution transmissivity
function of the average atmosphere above Paranal (provided by ESO) is
finely tweaked by scaling and changing the resolution until we can
exactly compensate for the observed absorption at various wavelengths
on the spectrum of the spectrophotometric standard; the coadded
spectrum of each galaxy is then corrected for the same absorption
pattern as for the closest (in time and airmass) standard star. We
have applied the telluric correction to both standard stars and
galaxies \textit{before} computing the sensitivity function and
applying the flux calibration.

In order to assemble a SED including the visible range and the H and K
NIR bands, absolute flux calibration is required. To this goal we have
obtained high SNR images in $H$ and $K_s$ bands in photometric
conditions with ISAAC as part of the same program. ``Total''
integrated magnitude derived from these images are used to rescale the
spectra. By ``total'' magnitude we adopt the magnitude integrated
within the petrosian aperture, as defined by the SDSS photometric
pipeline on the SDSS $r$-band images in data release 8
\citep[DR8,][]{SDSS_DR8}\footnote{Data are obtained from the DR8
  casjobs catalog server {\texttt
    http://skyservice.pha.jhu.edu/casjobs}.}. DR8 is chosen as it
fixes a number of issues especially with sky subtraction. $H$ and
$K_s$ magnitudes are used to rescale the respective ISAAC spectra in
the NIR, while the optical spectrum from the SDSS is normalised
according to the $r$-band petrosian magnitude\footnote{We adopt the
  petrosian magnitudes distributed in the SDSS-DR8. We have
  independently checked these values by doing our own aperture
  photometry on the SDSS images and found substantial agreement: our
  estimates are 0.011 mag fainter on average, with a scatter of 0.015
  mag. Only for two galaxies (PSB\,J1133$+$0205 and PSB\,J0151$-$0056)
  we find fainter magnitude by more than 0.03 mag (+0.040 and +0.033
  respectively). These small discrepancies are well within the errors
  adopted in Sec. \ref{sec:SSPcomp} and \ref{sec:CSPcomp}.}.
Conversions between NIR magnitudes natively calibrated in the Vega
system for the ISAAC filters into the corresponding flux
densities\footnote{Average flux densities within broad band filters
  are computed according to the integral defining AB magnitudes given
  by \cite{fukugita_etal96}.}  for our spectra are computed based on
the spectrum of Vega of \cite{kurucz_92}, distributed by BC03. Typical
total photometric uncertainties for ISAAC petrosian magnitudes are
estimated $\approx 0.05$ mag and include photometric zero point
fluctuations ($\approx 0.03$ mag, derived from repeated observations
of standard stars during the nights) and background subtraction
uncertainties.\footnote{Photometric measurements similarly performed
  on the UKIDSS \citep{lawrence+07} images that are available for a
  subsample of our galaxies are found in good agreement within the
  quoted uncertainties.} For the SDSS we assume errors of 0.02 mag,
which is the maximum reported for our objects by SDSS-DR8. Foreground
Galactic extinction correction is finally applied using the optical
depth derived from \cite{schlegel_dust} and assuming the standard
extinction laws of \cite{cardelli+89} and \cite{odonnell_94}.

\subsection{Aperture effects}\label{subsec:aper_eff}

The spectra of our galaxies, both SDSS and ISAAC, are taken in
relatively small apertures or slits centred on the brightness peak:
3-arcsec fibres are used in the SDSS with a typical seeing of $\approx
2$ arcsec while 1.5-arcsec long-slits with seeing of 1.4 arcsec or
better are used with ISAAC. Relative aperture size and seeing are
designed to conspire and collect a similar fraction of the galaxy
light. By rescaling the spectra according to the ``total'' petrosian
magnitudes we implicitly assume that stellar population gradients are
of limited relevance and the central region of the galaxies are
representative of the whole. \cite{pracy+12} have analysed a small
sample of post-starburst galaxies with integral field spectroscopy and
showed that, in reality, a large fraction of them have cores that are
younger than the rest of the galaxy, hence our assumption does not
strictly hold. Unfortunately, because of different seeing conditions
and geometry of the apertures, a precise and homogeneous rescaling of
the spectra is very hard to achieve. However, by comparing fluxes and
colours that are obtained in the petrosian and in the fibre apertures
we can show that the aperture effects for our sample are small enough
not to affect our conclusions.

First of all, we note that at the relatively high redshift of our
sample the apparent size of the galaxies is typically 2.5 arcsec
(median petrosian radius, maximum 3.1 arcsec), so that the 3-arcsec
SDSS fibre typically collects already half of the total flux. The
ISAAC spectra roughly collect the same fraction of flux. The long slit
geometry in this case integrates into the spectra not only the central
part, but also the outskirts of the galaxy, so that the ISAAC spectra
are indeed even more representative of the whole galaxy.

In second place, we have checked by how much ``total'' petrosian
colours differ from fibre colours in the optical. Fibre colours are
computed in such a way that they correspond as closely as possible to
the SED portion effectively captured in the SDSS spectra. The
strongest effects are expected and observed in $g-i$, since this
colour roughly corresponds to $u-r$ at $z\approx0.2$ and therefore is
the most sensitive to stellar population properties by bracketing the
D4000 break. We find that, on average, fibre $g-i$ is bluer by 0.027
mag with respect to the corresponding total petrosian colours, with a
rms of 0.037. Only four galaxies (out of 16) have a fibre colour
which is bluer than total by more than 0.05 mag: PSB\,J1046$+$0714,
PSB\,J1015$+$0103, PSB\,J0227$-$0015 and PSB\,J0328$+$0045. We dub
these galaxies ``blue core'' in the following. For the remainder,
differences in colour can be considered as not significant, although
the data hint at starbursts preferentially occurring in the central
regions of galaxies rather than in the outskirts, consistently with
\cite{pracy+12}. Looking at $r-i$, which roughly corresponds to the
rest-frame optical colour $\mathrm{m_{0.55}}-\mathrm{m_{0.70}}$ that
will be discussed in the following, we find even smaller differences
of 0.012 mag on average (0.021 mag rms). These are well within the
spectrophotometric calibration errors.

Finally, we estimate the relative aperture corrections between the
optical and NIR bands. Specifically, we consider the aperture
corrections in $r$ and $H$-band, defined by
$ap\_corr=m_{\mathrm{fiber}}-m_{\mathrm{petro}}$. For the $r$-band we
use the quantities provided by the SDSS database. For the $H$-band we
use the total petrosian magnitude as derived above, while to obtain
fibre magnitudes we first convolve the ISAAC images to a common PSF of
FWHM$=$2 arcsec and then integrate the flux within a circular aperture
of 3 arcsecond in diameter, as done in the SDSS pipeline. Since for
most of the ISAAC images we lack a sufficient number of high SNR stars
to compute an accurate PSF, the final FWHM is approximated within 0.1
arcsec. We find an average difference of aperture corrections in $r$
and $H$-band $<ap\_corr_r-ap\_corr_H>=0.026$ mag with rms of 0.1 mag,
indicating that on average our galaxies are slightly more concentrated
in $r$ than in $H$, although with very marginal significance. This is
consistent with the existence of few blue cores and, in general, of
very small systematic SED variations between the spectral and the total
apertures.

This analysis of the aperture effects justifies our choice to use the
total petrosian magnitude to compute the relative normalisation of the
optical and NIR spectra. Total petrosian magnitudes are much more
stable than magnitudes integrated in small apertures comparable with
the seeing, which therefore rely on uncertain PSF convolution. On the
other hand, we have shown that they do not introduce significant
biases between different spectral regions. The systematic aperture
bias of 0.026 mag between $r$ and $H$ is in fact well within the
photometric uncertainties estimated in the previous section.

In the following analysis we will adopt 0.1 mag as standard error on
the optical-NIR colours and 0.05 mag on the optical colours,
to take into account both photometric errors and aperture effects.

\section{Results and Analysis}\label{sec:analysis}
\subsection{Composite optical-NIR spectra: absence of sharp NIR
  features}\label{sec:noNIRfeatures}
The composite optical-NIR spectra obtained as explained in the
previous section are presented in Fig. \ref{fig:specs_full}, in the
rest frame and normalised to the flux density at 5500~\AA, in order to
ease a prompt comparison between all galaxies. The actual
normalisation refers to the median flux in the range 5480--5520~\AA,
in order to minimise the effect of noise on the individual spectral
pixels. The SDSS portion of the spectra covers approximately the
rest-frame range 3000--7500~\AA, ISAAC H 1.25--1.50~$\mu$m, and ISAAC
K 1.65--2.00~$\mu$m. For each galaxy, three panels display the full
optical-NIR spectral range (top left plot), a zoom into the optical
region that includes the most significant diagnostic lines (top right
plot), and the NIR H--K range (bottom plot). Green lines are used to
plot the spectra at high resolution: the original $\approx 40$ and
$50$~\AA~ resolution at $\approx 4$ and $6$~\AA~ per pixel in the rest
frame is used for the H and K spectra respectively, while for the SDSS
spectra (original resolution $2.5$~\AA~ at $\approx 1$~\AA~ per pixel
in the rest frame) a smoothing over 3 pixels is applied to make the
plot more readable. Black thick lines are used to (over)plot heavily
smoothed versions of the spectra, in order to ease the identification
of possible spectral features, especially in the NIR bands: smoothing
over 15 pixels is adopted both in the optical and the NIR spectra,
resulting in rest frame resolution of $\approx 15$, 60 and 90~\AA~ in
optical, H and K ranges, respectively. Vertical dotted lines mark the
position of the main (expected) spectral features: the Balmer series
(up to H$\eta$ 9-2) and the two features of CN (1.41~$\mu$m) and C$_2$
(1.77~$\mu$m) due to carbon rich AGB stars
\citep{lancon_mouhcine_02}. For each galaxy in the bottom panel we
report also the light-weighted stellar age and metallicity and
corresponding uncertainties derived from optical absorption features
as in \cite{gallazzi+05}.
\begin{figure*}
\centerline{
\includegraphics[width=0.49\textwidth]{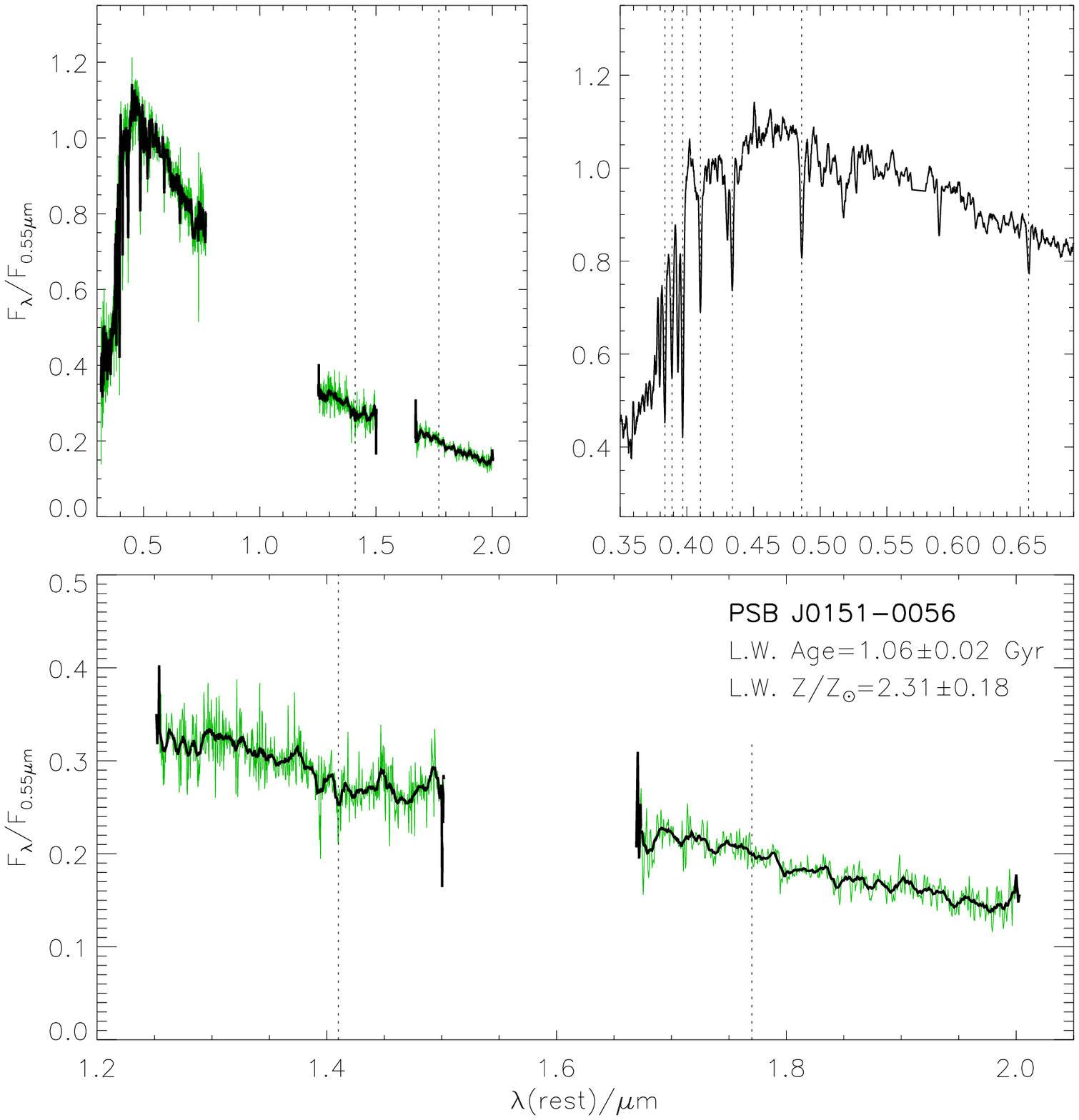}
\includegraphics[width=0.49\textwidth]{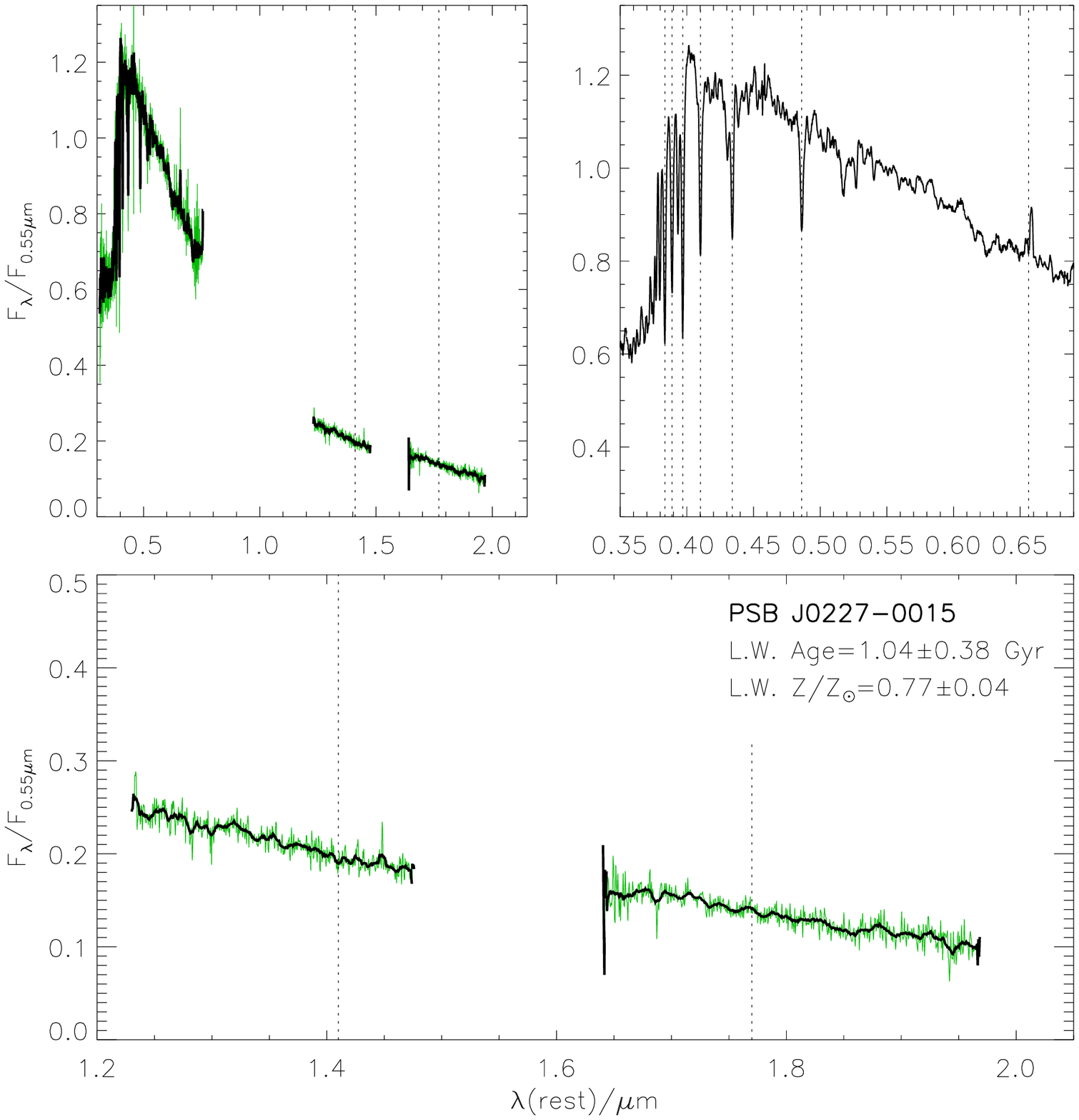}
}
\centerline{
\includegraphics[width=0.49\textwidth]{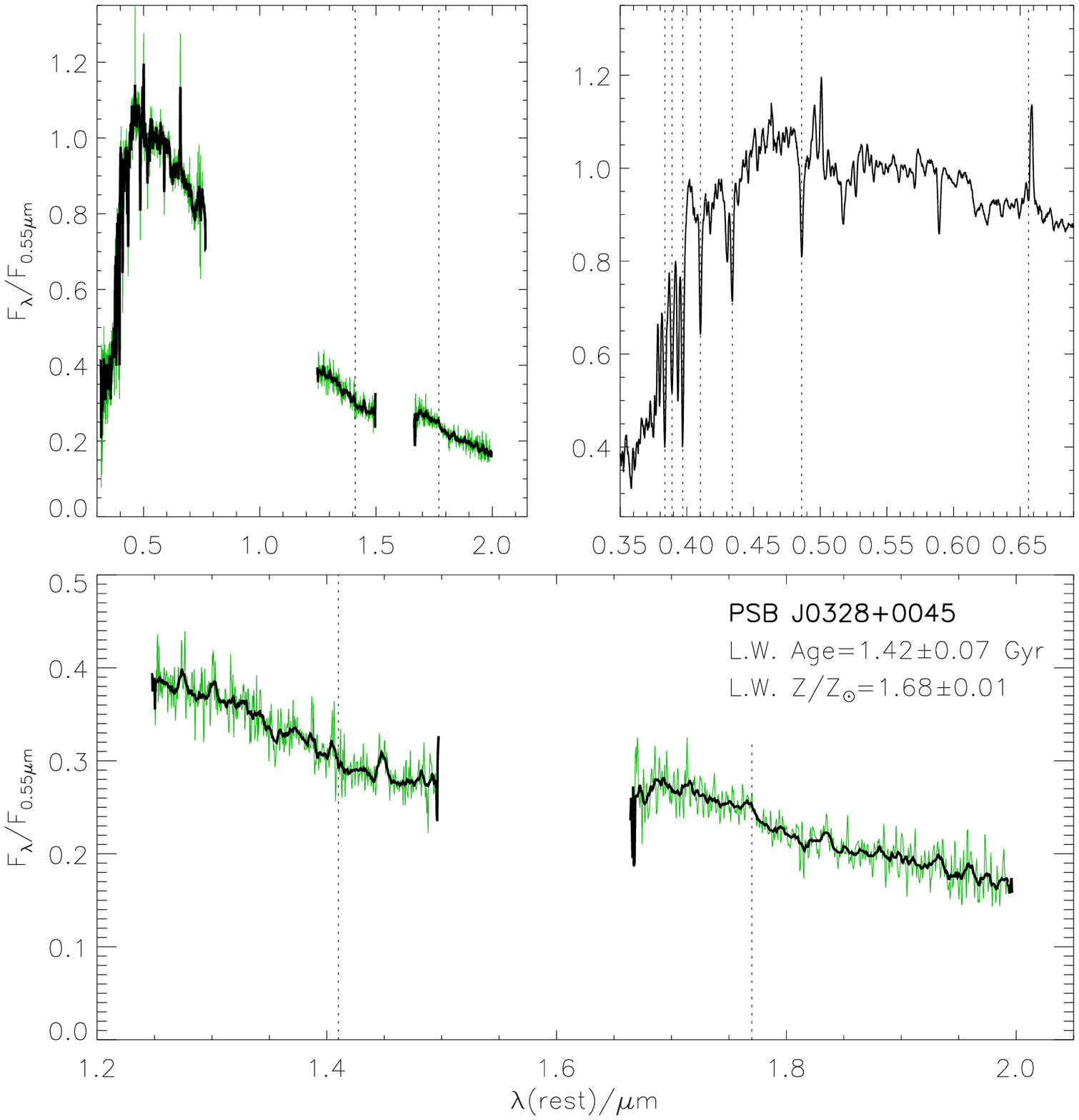}
\includegraphics[width=0.49\textwidth]{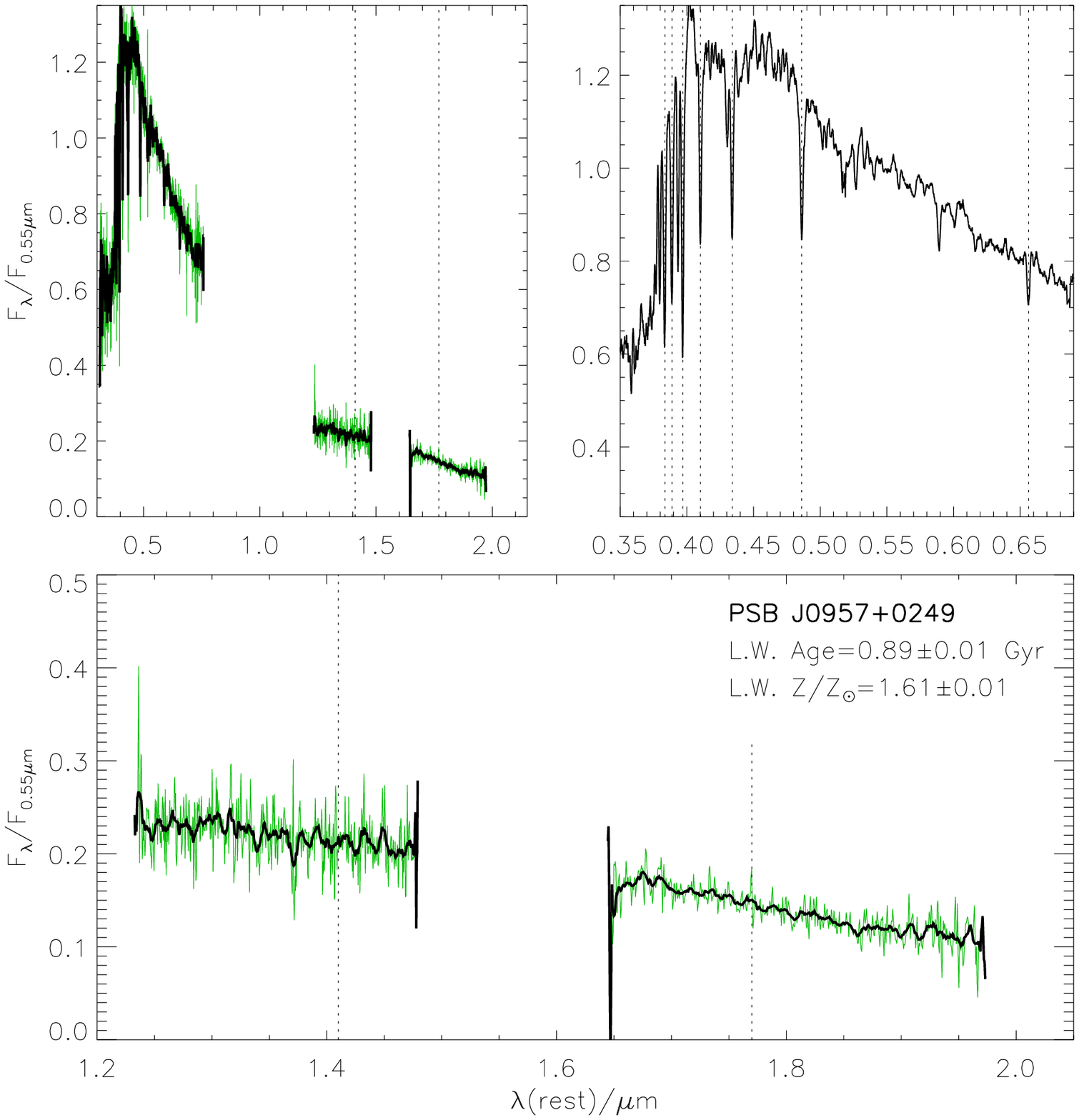}
}
\caption{Composite rest-frame spectra for the sixteen galaxies of the
  sample, including optical SDSS spectra and the new NIR H and K band
  spectra observed with ISAAC at the ESO-VLT. Correction for
  foreground Galactic extinction and normalisation at 5500~\AA~are
  applied after relative normalisation between the optical and the NIR
  spectra based on SDSS $r$-band photometry and our own ISAAC imaging,
  as detailed in the text. The full/high resolution version of the
  spectra is plotted in green (rest-frame resolution 2.5~\AA~for the
  SDSS and 40 and 50~\AA~for ISAAC); the thick black curves are
  smoothed versions of the same spectra, to filter out the noise and
  improve the visibility of the features. The resulting effective
  resolution in these smoothed spectra is $\approx 15$, 60 and
  90~\AA~rest-frame in optical, H and K ranges, respectively.  The new
  NIR H and K band spectra observed with ISAAC at the ESO-VLT are
  displayed in the bottom plot of each panel. The full optical-NIR
  range is shown in the top-left plot, while a zoom-in of the optical
  range is shown in the top-right plot. Vertical dotted lines mark the
  position of the main (expected) spectral features: the Balmer series
  (up to H$\eta$ 9-2) and the two features of CN (1.41~$\mu$m) and
  C$_2$ (1.77~$\mu$m) due to carbon rich AGB stars
  \citep{lancon_mouhcine_02}. Light-weighted ages and metallicity are
  reported for each galaxy (see text for
  details).}\label{fig:specs_full}
\end{figure*}
\addtocounter{figure}{-1}
\begin{figure*}
\centerline{
\includegraphics[width=0.49\textwidth]{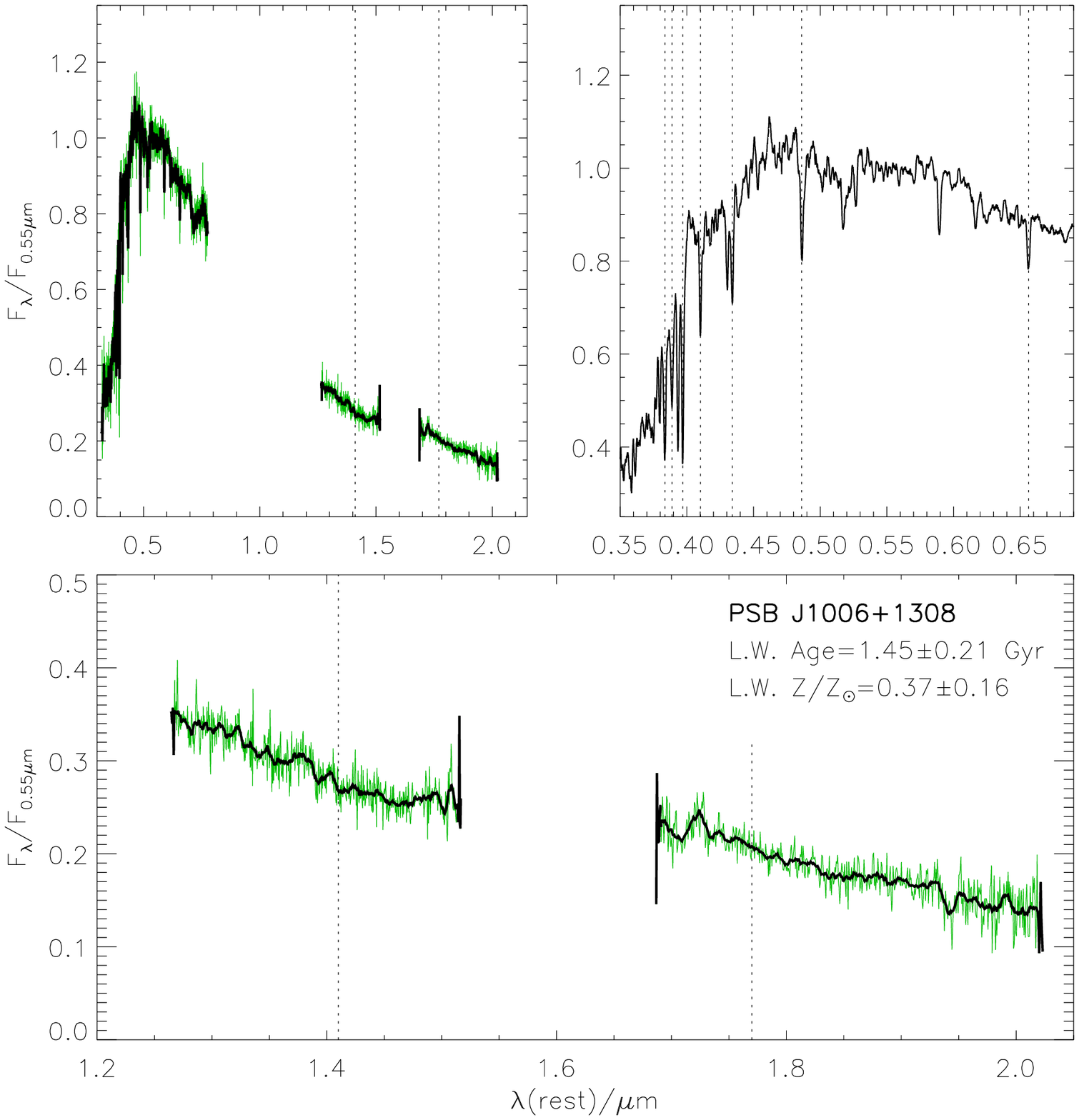}
\includegraphics[width=0.49\textwidth]{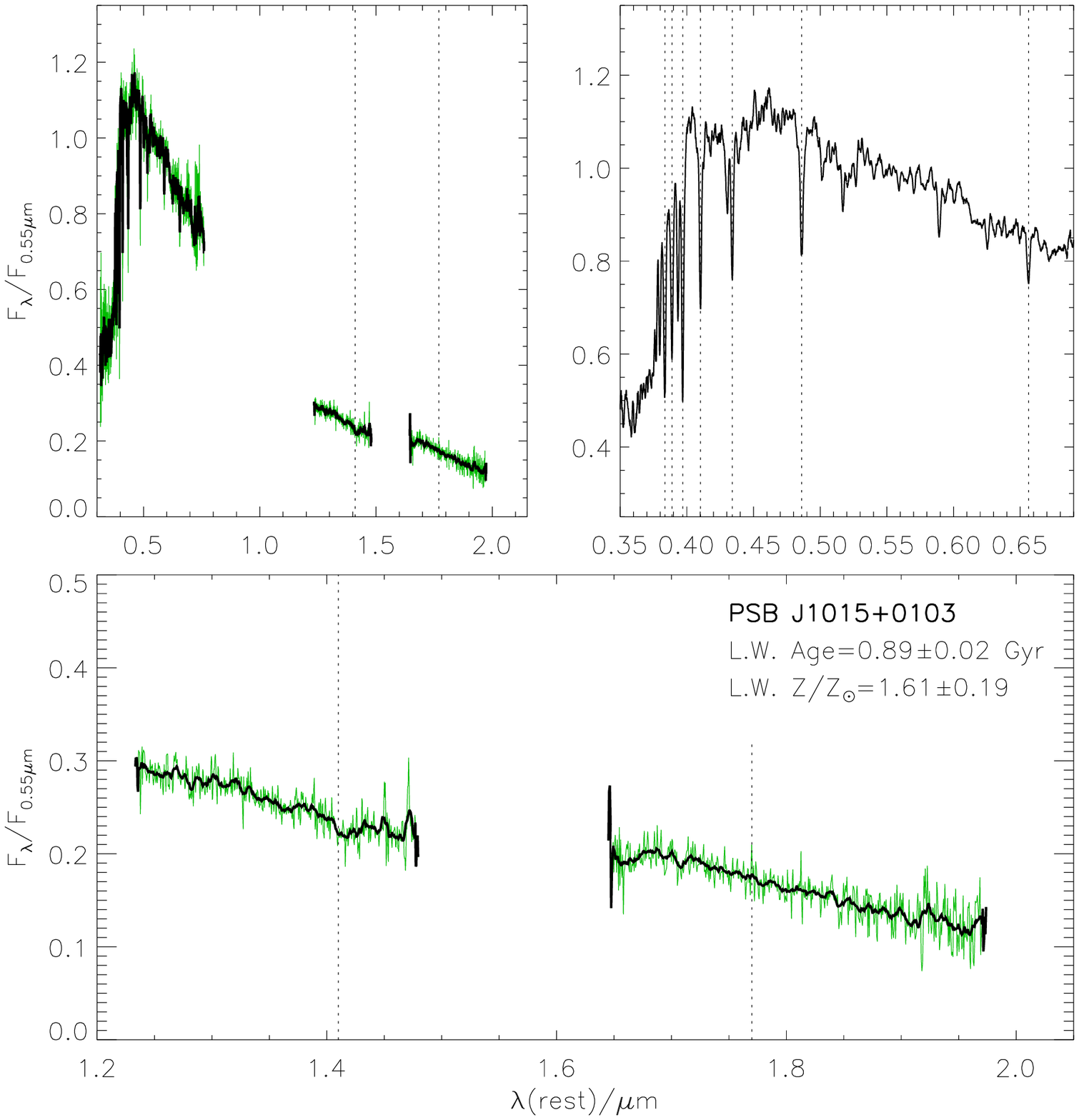}
}
\centerline{
\includegraphics[width=0.49\textwidth]{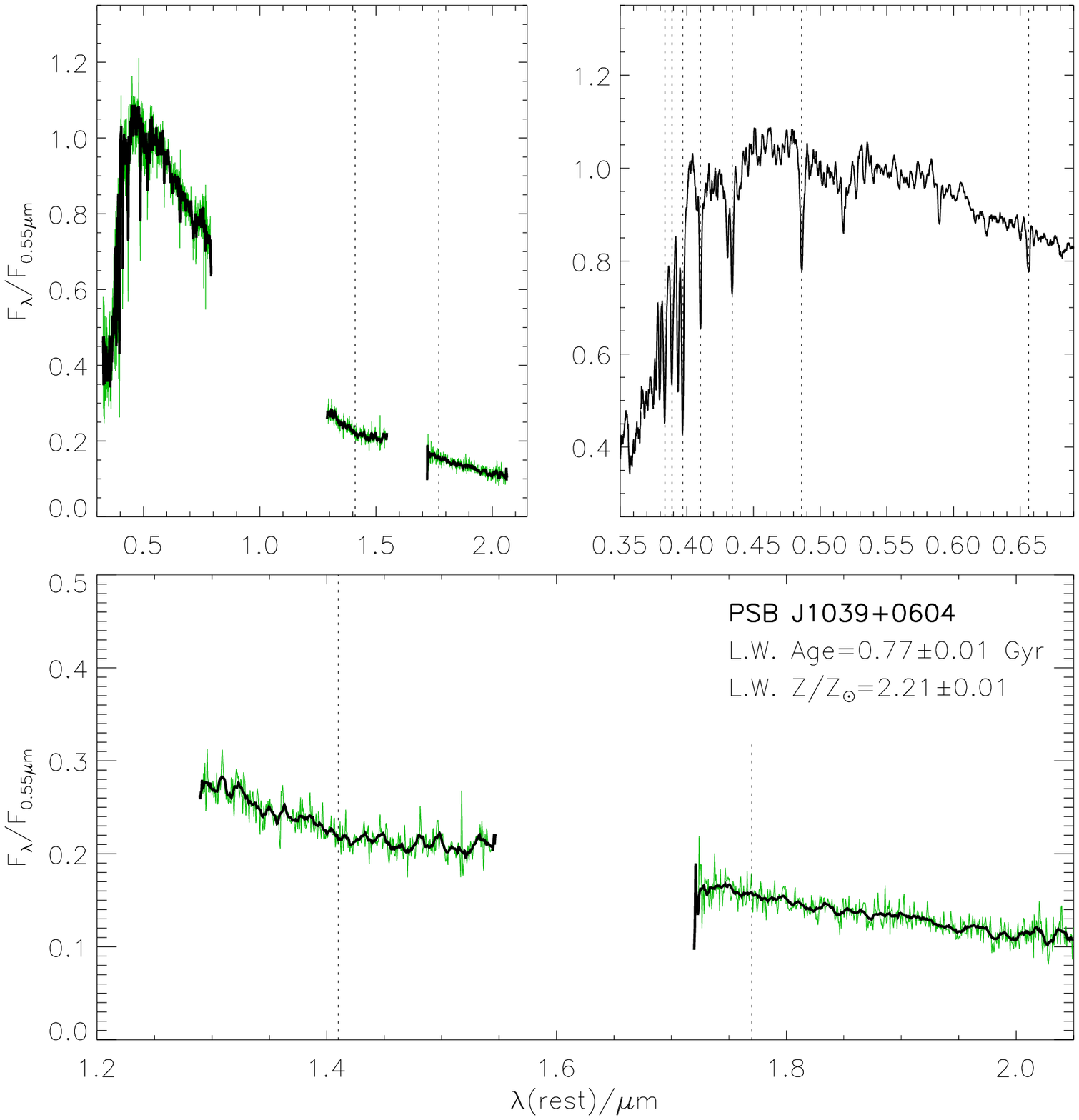}
\includegraphics[width=0.49\textwidth]{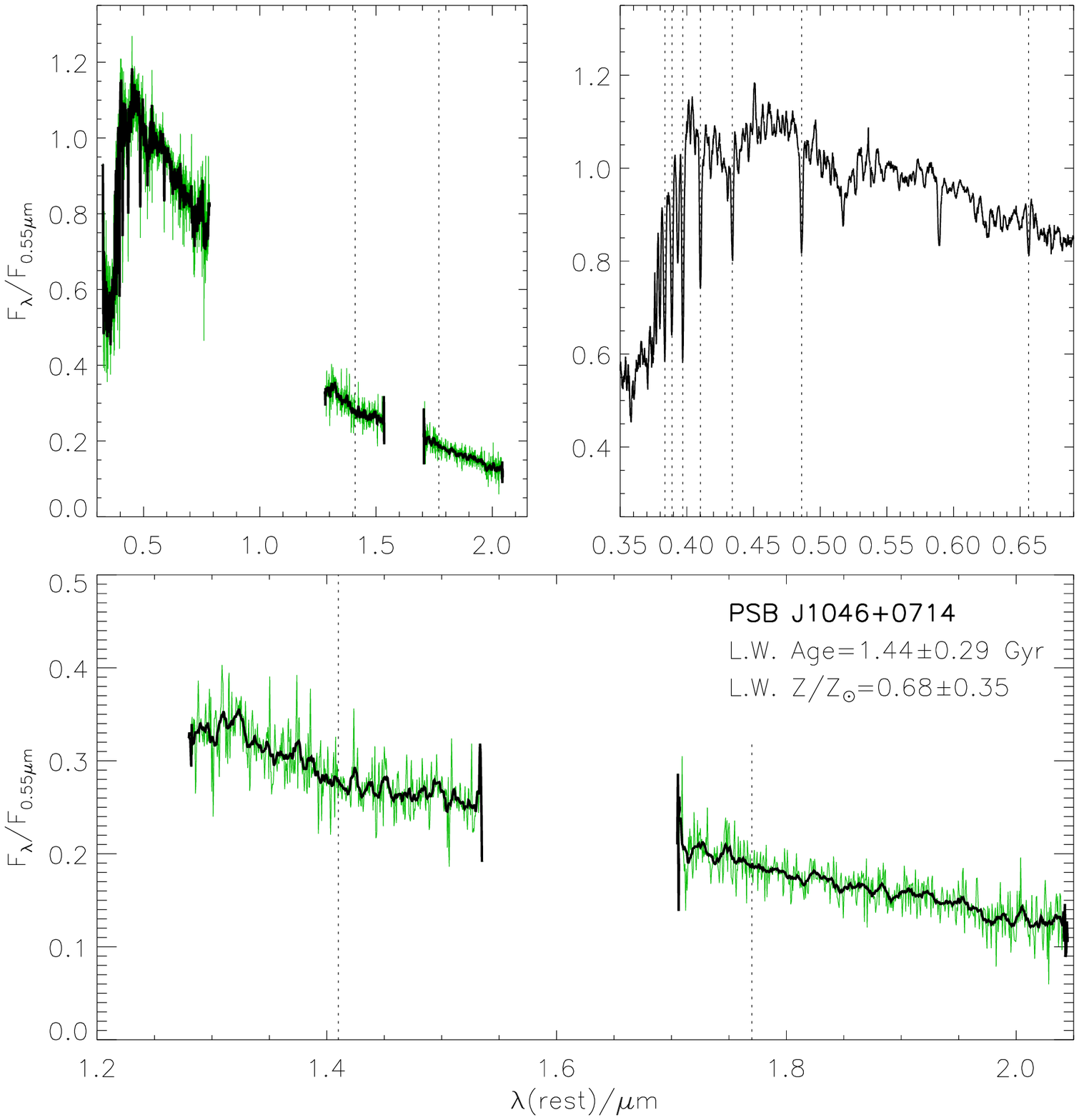}
}
\caption{\bf (continued)}
\end{figure*}
\addtocounter{figure}{-1}
\begin{figure*}
\centerline{
\includegraphics[width=0.49\textwidth]{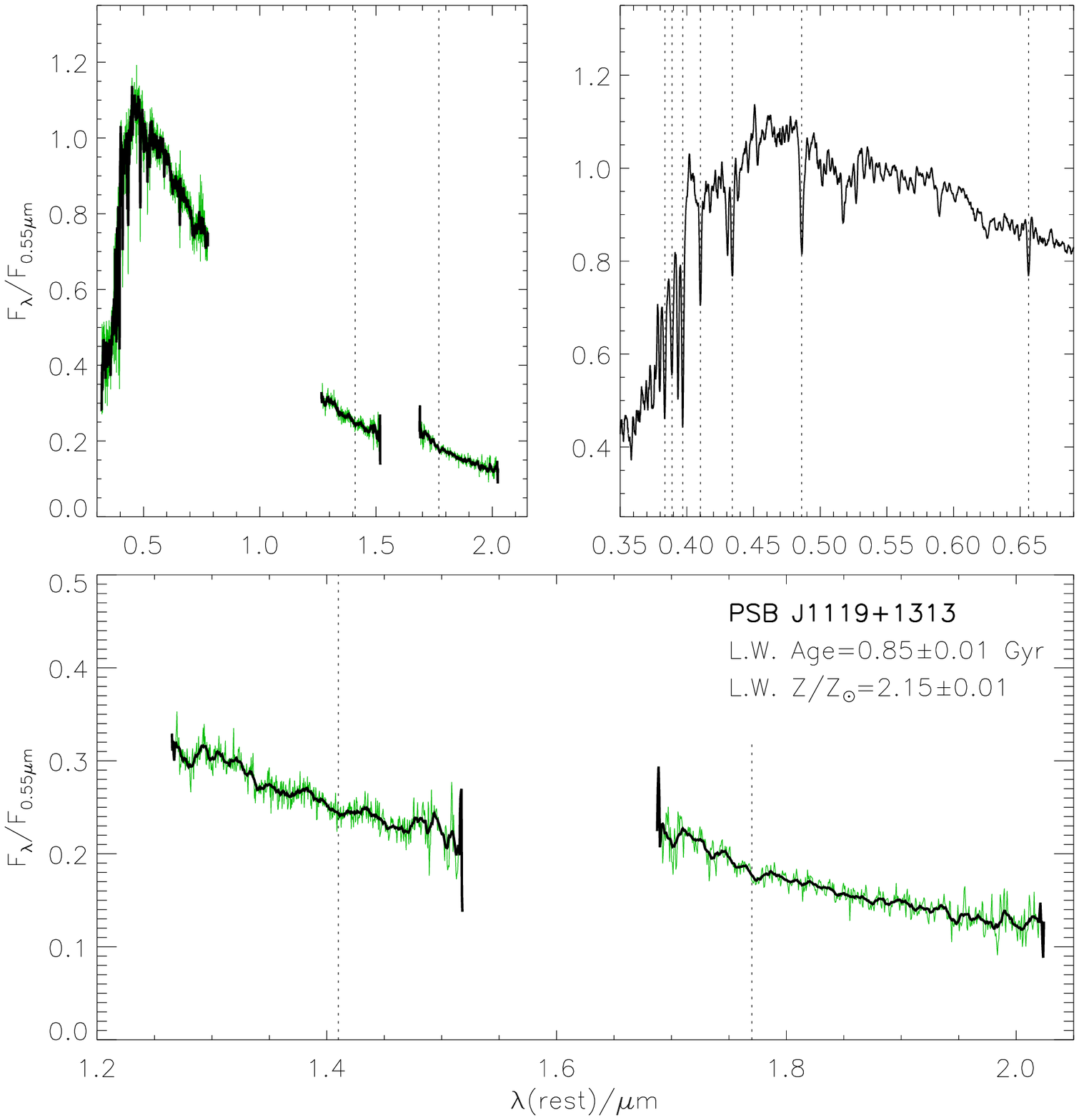}
\includegraphics[width=0.49\textwidth]{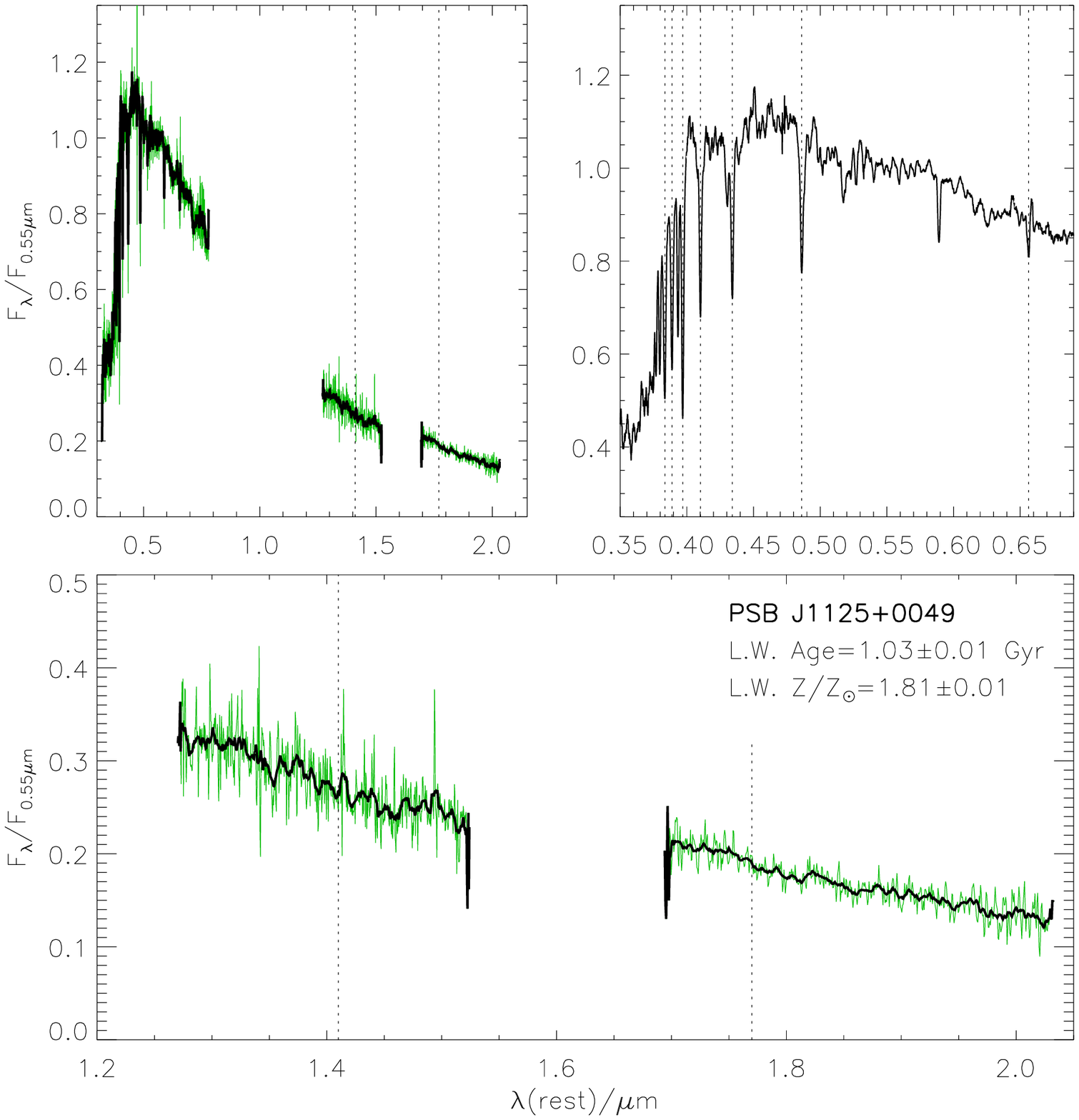}
}
\centerline{
\includegraphics[width=0.49\textwidth]{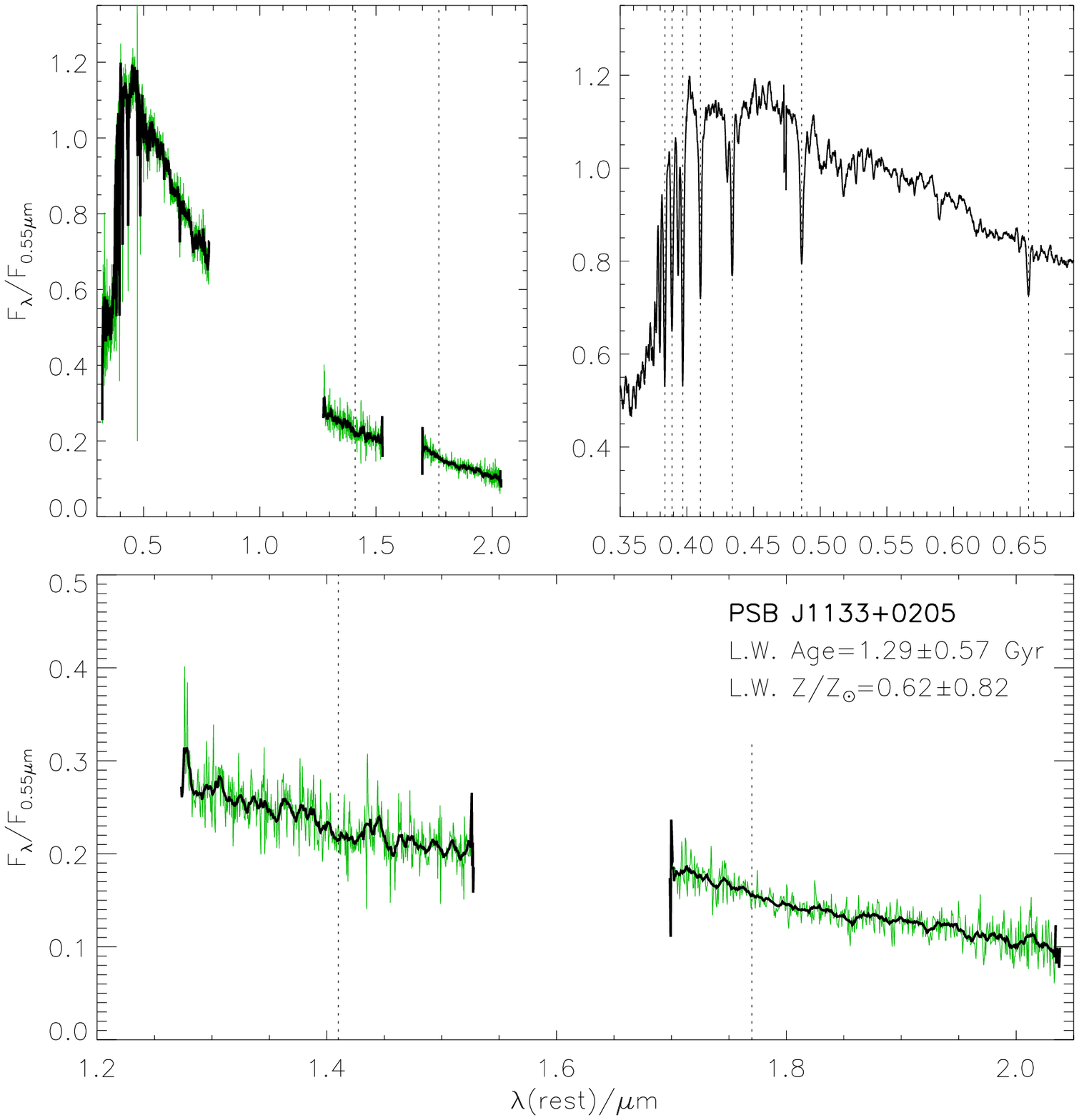}
\includegraphics[width=0.49\textwidth]{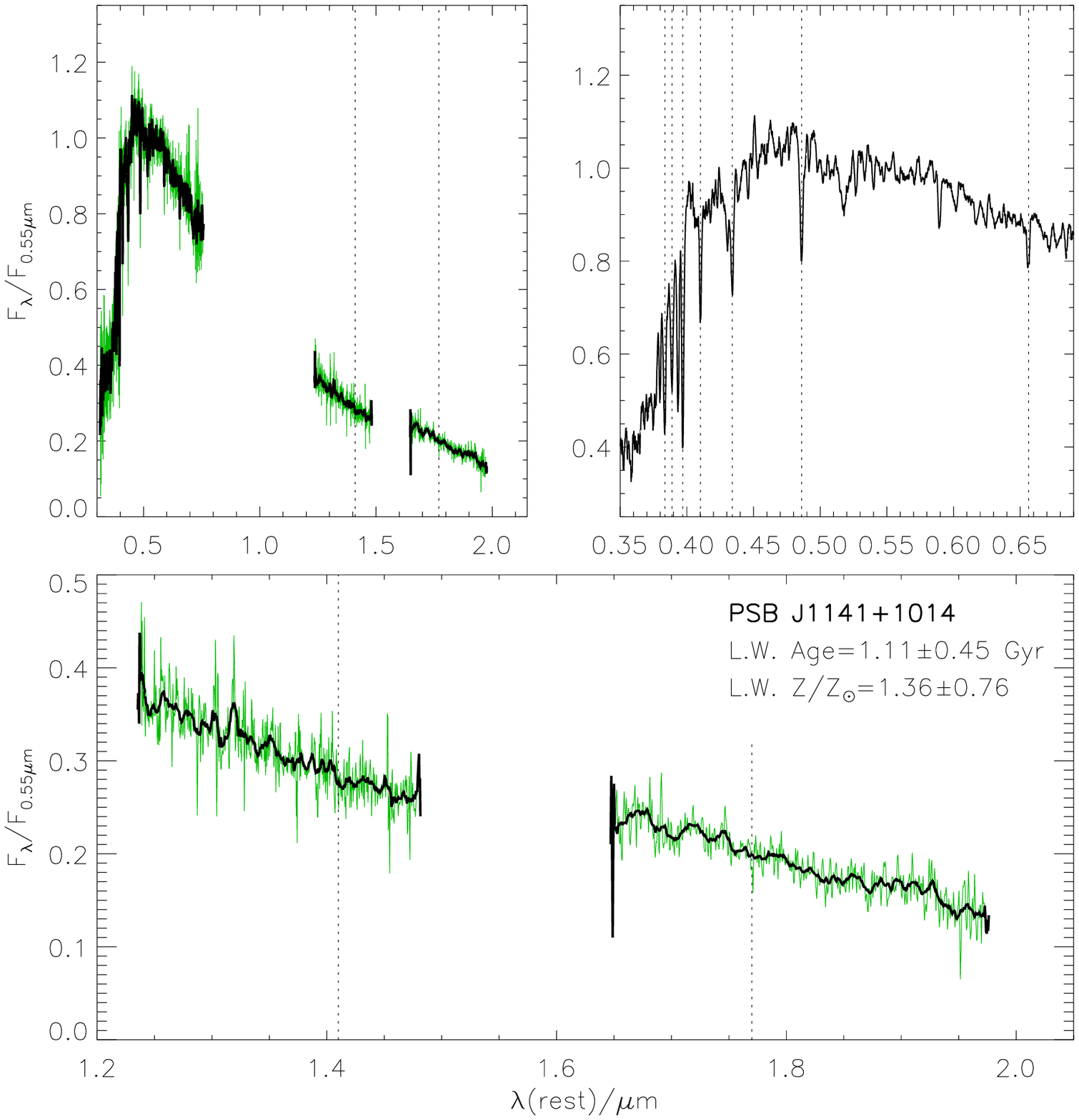}
}
\caption{\bf (continued)}
\end{figure*}
\addtocounter{figure}{-1}
\begin{figure*}
\centerline{
\includegraphics[width=0.49\textwidth]{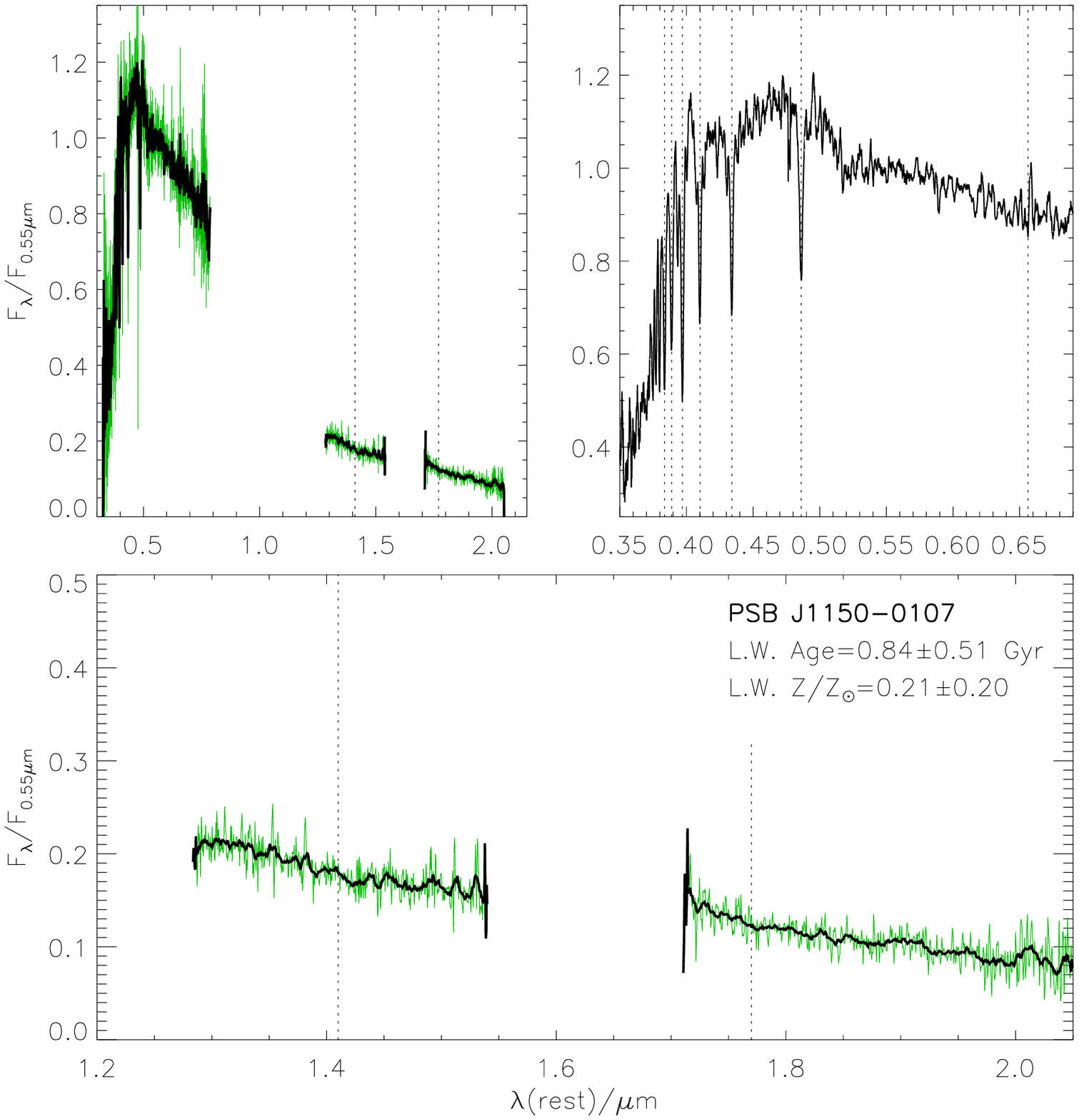}
\includegraphics[width=0.49\textwidth]{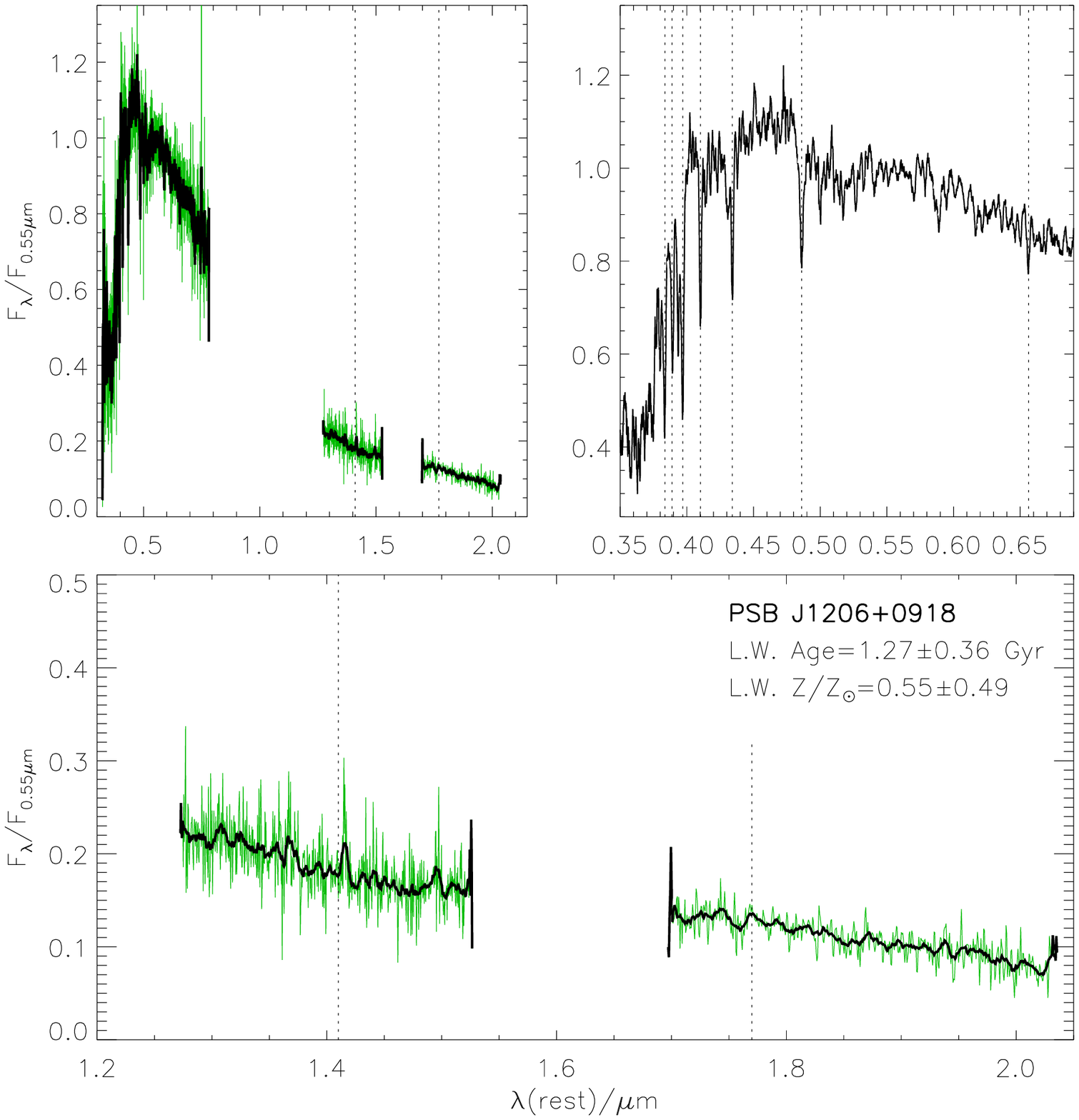}
}
\centerline{
\includegraphics[width=0.49\textwidth]{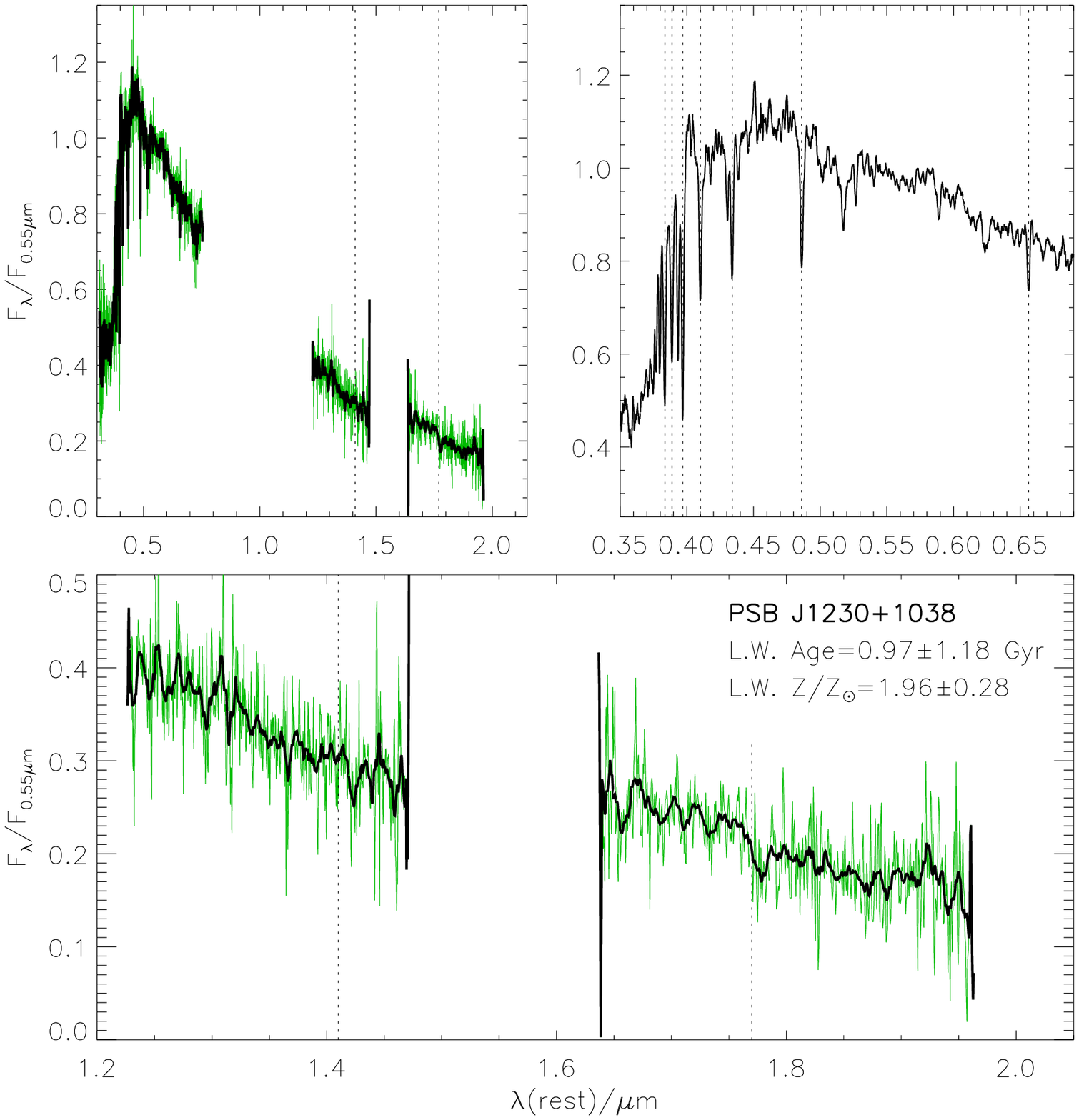}
\includegraphics[width=0.49\textwidth]{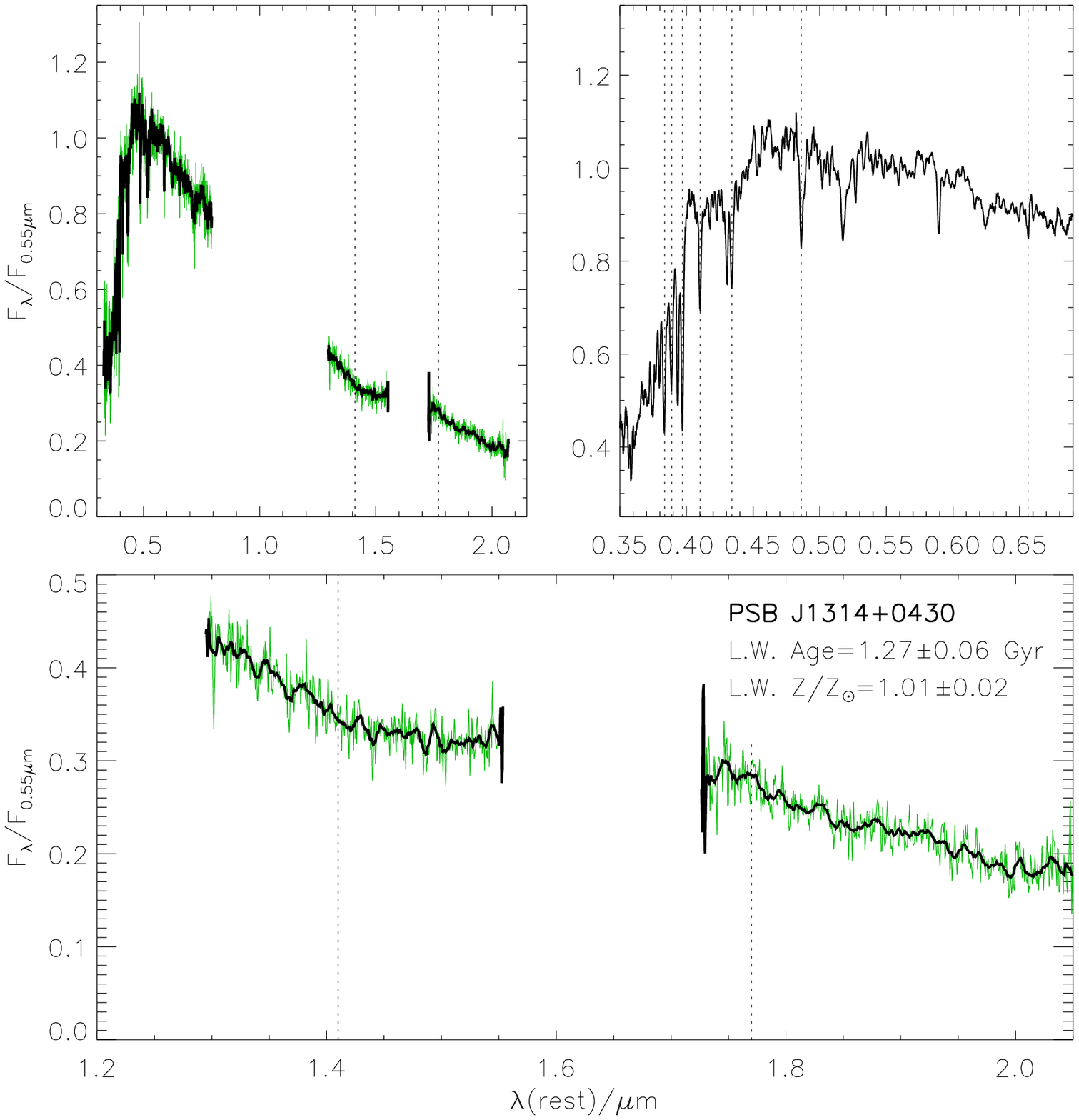}
}
\caption{\bf (continued)}
\end{figure*}

All galaxies display typical E+A or post-starburst spectra with very
strong Balmer absorptions, well developed D4000 break and blue
continuum. The actual slope of the optical continuum generally shows
the well known dependence on age and metallicity, although dust
attenuation may affect the slope as well. We do not attempt to put any
constraint on this parameter, but discuss its possible impact on our
conclusions below. Three galaxies, PSB\,J0227$-$0015,
PSB\,J0328$-$0045, and PSB\,J1150$-$0107 show significant H$\alpha$
emission (and some higher ionisation lines for PSB\,J0328$-$0045),
indicating some residual ongoing star formation, which is also
confirmed by (marginal) detections at 22~$\mu$m (W4) in the WISE All
Sky Survey \citep{wright+10}, measured with SNR of 4.4 and 2.8, and
2.0, respectively. However, the equivalent width in all cases matches
the \cite{goto_05} criteria, being $-2.27\pm0.10$~\AA,
$-2.80\pm0.10$~\AA, and $-2.43\pm0.24$~\AA, respectively, all larger
than $-3$~\AA. Given the observed strength of all Balmer absorption
lines, the contribution of newly formed stars (or AGN) can be largely
neglected in comparison to the radiation by the intermediate-age
population we aim at studying here. It is worth noting that two of
these galaxies (PSB\,J0227$-$0015 and PSB\,J0328$-$0045) also have a
blue core, supporting the idea that blue cores might be related to
some residual star formation extending close to the present time. The
absence of emission lines and blue cores from the majority of the
galaxies in our sample further confirms their nature of genuine
post-starburst galaxies.

What is most important for the scope of this work is the lack of any
evidence for sharp features in the NIR in any of the 16 galaxies. Most
of them display completely featureless continua with approximately
uniform slope within the two NIR windows. Some tentative evidence for
a wavelength-dependent slope is seen in a few objects:
PSB\,J1039$+$0604, PSB\,J1046$+$0714 and especially PSB\,J1314$+$0430
appear to change their slope around 1.41~$\mu$m, while in
PSB\,J1119$+$1313 and PSB\,J1133$+$0205 a change of slope can be
tentatively seen around 1.77~$\mu$m. Similar mild changes of slope,
most likely due to molecular bands in the atmospheres of cool stars,
are compatible with BC03 but are in fact substantially dissimilar from
the sharp drops by a few 10\% in flux predicted by Ma05 models (see
Fig. \ref{fig:Ma05vsBC03}).

\begin{figure*}
\centerline{
\includegraphics[width=\textwidth]{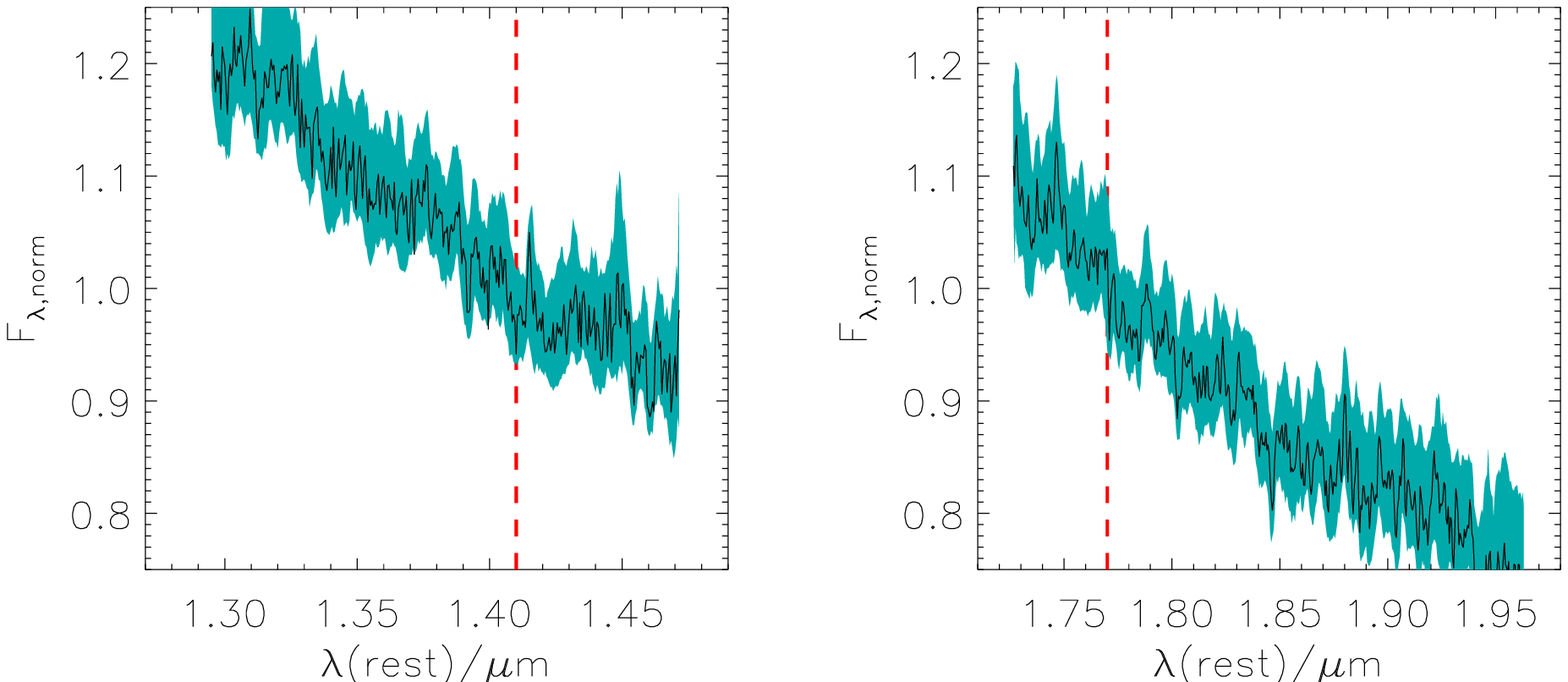}
}
\caption{Stacked H (\textit{left panel}) and K (\textit{right panel})
  spectra of the 16 galaxies. Each spectrum has been re-normalized to
  1 in a wavelength range of 1000~\AA~around the expected positions of
  the features at 1.41 and 1.77~$\mu$m (marked with red dashed vertical
  lines), respectively. Band-head drops of more than a few per cents
  \emph{on average} can be ruled with very high
  confidence.}\label{fig:stack_HK}
\end{figure*}

To further support these conclusions, we stacked all 16 spectra in H
and K separately, after normalising them in a 1000~\AA-wide region
around 1.41 and 1.77~$\mu$m, respectively. The results are displayed in
Fig. \ref{fig:stack_HK}, with the black lines showing the average
spectrum and the shaded green area indicating the 16-84\%
inter-percentile range (approximately $\pm 1\sigma$ range). This
figure shows that \emph{on average} carbon molecules in the
atmospheres of evolved stars produce band-head drops that are
definitely less than 5\% of the continuum.

\subsection{Optical-NIR colours vs optical absorption indices}\label{sec:SSPcomp}

Stellar absorption indices in the optical range are extremely powerful
diagnostics of stellar population properties. Decades of development
and tests have led to substantial agreement on their interpretation in
terms of ages and global metallicity. In this section we exploit part
of the wealth of information that we can extract from the optical SDSS
spectra of our galaxies for two main purposes: \textit{i)} probe the
ability of ``TP-AGB light'' models (represented by BC03) and ``TP-AGB
heavy'' models (Ma05) to reproduce the observed trends and
distributions in the parameter space of optical absorption indices and
optical-NIR colours; \textit{ii)} investigate whether the lack of NIR
spectral features can be due to selection biases such that we
misinterpreted optical diagnostics (e.g. due to the use of BC03-based
model libraries) and ended up missing the relevant phase for TP-AGB
stars.

\begin{figure*}
\centerline{
\includegraphics[width=\textwidth]{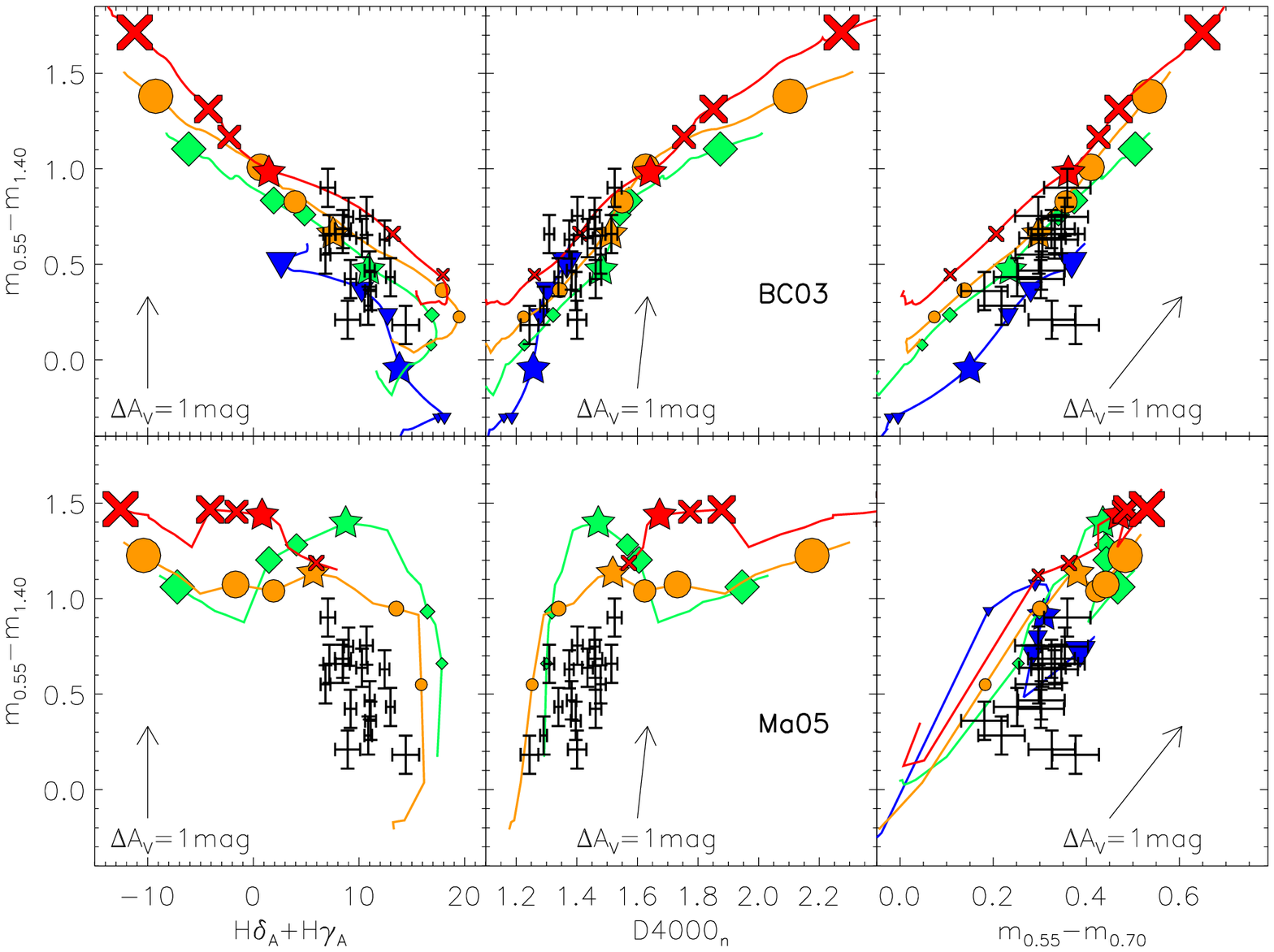}
}
\caption{Comparison between the observed properties of the
  post-starburst galaxies in our sample (points with error bars) and a
  suite of SSP models (coloured symbols and lines) from BC03
  (\emph{top row}) and Ma05 (\emph{bottom row}). The optical-NIR
  colour $\mathrm{m_{0.55}}-\mathrm{m_{1.40}}$ is plotted against
  H$\delta_{\mathrm A}+$H$\gamma_{\mathrm A}$ (\emph{left panels}),
  D4000$_{\mathrm n}$ (\emph{central panels}), and the optical colour
  $\mathrm{m_{0.55}}-\mathrm{m_{0.70}}$ (\emph{right panels}). Each
  track (blue with upside-down triangles, green with diamonds, orange
  with circles and red with crosses) corresponds to metallicity of
  1/50, 0.4, 1 and 2.5 solar, respectively, for the BC03 models in
  \textit{top row}, and to metallicity of 1/20, 0.5, 1 and 2 solar,
  respectively, for the Ma05 models in \textit{bottom row}. The tracks
  for BC03 and the colour-colour plot of Ma05 extend from 50 Myr to
  13.6 Gyr. As Ma05 only provides low-resolution SEDs,
  H$\delta_{\mathrm A}+$H$\gamma_{\mathrm A}$ and D4000$_{\mathrm n}$
  for these models are taken from the corresponding STELIB-based SSPs
  published by Maraston \& Str\"omb\"ack (2011), limited to three
  metallicities, 0.5, 1.0 and 2.0 solar, and starting with ages of
  200, 30, 400 Myr, respectively. Stars mark 1 Gyr, while symbols of
  increasing size mark ages of 0.3, 0.5, 1.5, 2 and 10 Gyr. IMF is
  assumed Salpeter in all cases. The arrow in each plot shows the
  shift produced by $A_V=1$ mag for a uniform attenuation following
  $\tau_\lambda\propto\lambda^{-0.7}$. While colours and spectral
  indices can be reproduced simultaneously by BC03 SSPs, Ma05 SSPs
  yield optical-NIR colours that are inconsistent with age-sensitive
  spectral indices and optical colours, unless metallicities as low as
  1/20 solar are advocated for all
  galaxies.}\label{fig:diag_optNIR_SSP}
\end{figure*}

We start considering only age-sensitive absorption indices and
optical-NIR colours and build the diagnostic plots presented in
Fig. \ref{fig:diag_optNIR_SSP} for BC03 and Ma05 in the \emph{top row}
and \emph{bottom row}, respectively: coloured symbols and lines show
the predicted optical-NIR colour $\mathrm{m_{0.55}}-\mathrm{m_{1.40}}$
as a function of $\mathrm{H\delta_A+H\gamma_A}$, $\mathrm{D4000_n}$,
and the optical colour $\mathrm{m_{0.55}}-\mathrm{m_{0.70}}$ (from
\emph{left} to the \emph{right}).\footnote{Here $\mathrm{m_{0.55}}$,
  $\mathrm{m_{0.70}}$ and $\mathrm{m_{1.40}}$ denote the AB magnitudes
  corresponding to the flux density around 0.55, 0.70 and 1.40~$\mu$m
  (rest), respectively.} For BC03, ages range from 50 Myr (lower ages
are not relevant for the present analysis as they would yield much
bluer colours than observed) to 13.6 Gyr. Metallicities range from
1/50 solar to 2.5 times solar, and are represented with different
symbols and colours from blue to red (solar in orange, see figure
caption). Symbols of increasing size are used to mark discrete
increasing ages, with the crucial age of 1 Gyr identified by a
star. The spectral indices $\mathrm{H\delta_A+H\gamma_A}$ and
$\mathrm{D4000_n}$ are measured after convolving the model spectra to
match the resolution and average velocity dispersion of the galaxies
($<\sigma>=164\mathrm{km~s^{-1}}$). The Ma05 SSP spectra are
distributed over the full optical-NIR range only at low resolution
(20~\AA~sampling), therefore cannot be used to compute reliable
indices, $\mathrm{H\delta_A+H\gamma_A}$ in particular.\footnote{The
  effect of such a coarse sampling is to underestimate
  $\mathrm{H\delta_A+H\gamma_A}$ by up to $\approx 5$~\AA~and to
  overestimate $\mathrm{D4000_n}$ by up to $\approx 0.1$.}
\cite{maraston_stroembaeck_11} delivered high resolution spectra for a
subsample of the Ma05 age-metallicity grid, over the optical range:
only three metallicities are available (0.5, 1.0 and 2.0 solar), with
the lowest ages being 200, 30, 400 Myr, respectively. We use these
models (in their STELIB version), convolved to the effective
resolution of our data, to compute the $\mathrm{H\delta_A+H\gamma_A}$
and $\mathrm{D4000_n}$ plotted in the first two panels of the
\emph{bottom row} of Fig. \ref{fig:diag_optNIR_SSP}; optical-NIR
colours are computed from the original Ma05 SEDs of corresponding age
and metallicity. For the colour-colour plot the full grid of Ma05 is
used instead, which allows extending the metallicity down to 1/20
solar and the age down to 50 Myr as for BC03.  Observational data
points derived from our composite optical-NIR spectra are plotted as
error bars.\footnote{Errors on the colours are uniformly set equal to
  0.05 and 0.1 mag for the optical colour
  $\mathrm{m_{0.55}}-\mathrm{m_{0.70}}$ and the optical-NIR colour
  $\mathrm{m_{0.55}}-\mathrm{m_{1.40}}$ respectively, following the
  discussion in Sec.  \ref{subsec:aper_eff}. Errors on indices are
  derived from the MPA-JHU catalog and, in the case of
  $\mathrm{D4000_n}$, incremented \textit{a posteriori} to take into
  account flux calibration errors, as statistically derived by
  comparing duplicate observations of a large sample of SDSS
  galaxies.}

As shown by the \textit{top} row of Fig. \ref{fig:diag_optNIR_SSP},
BC03 SSPs provide a good match to the observed spectral diagnostics,
with the only exception of a couple of galaxies, PSB\,J1150$-$0107 and
PSB\,J1206$+$0918, which appear to have a too blue
$\mathrm{m_{0.55}}-\mathrm{m_{1.40}}$ NIR-optical colour for the given
$\mathrm{m_{0.55}}-\mathrm{m_{0.70}}$ optical colour (or, reversely, a
too red $\mathrm{m_{0.55}}-\mathrm{m_{0.70}}$ for the given
$\mathrm{m_{0.55}}-\mathrm{m_{1.40}}$) and will be discussed in
Sec. \ref{subsec:deviant_objs}. In particular, galaxies cover the
range in age and metallicity corresponding to light-weighted ages and
metallicities listed in Table \ref{tab:sample}, which are mostly
relevant for the TP-AGB phase.

As far as the comparison with the Ma05 models is concerned, we note
that the observed age-sensitive indices approximately match the range
of ages given by our light-weighted estimates. Most of the observed
$\mathrm{H\delta_A+H\gamma_A}$ and $\mathrm{D4000_n}$ values
correspond to SSP ages between 0.5 and 1 (1.5) Gyr for the solar
(sub-solar) metallicity track, while ages smaller than 0.5 Gyr would
be inferred from the 2 times solar metallicity track. For the given
indices, however, the predicted $\mathrm{m_{0.55}}-\mathrm{m_{1.40}}$
NIR-optical colours are on average 0.5 mag redder than the observed
ones. This is consistent with the Ma05 ``TP-AGB heavy'' models having
the NIR very much boosted with respect to BC03 in the age range around
1 Gyr implied by the Balmer indices. In the rightmost panel of
Fig. \ref{fig:diag_optNIR_SSP}, \textit{bottom} row, we see that,
unless a metallicity as low as 1/20 solar is representative of most
galaxies and ages are significantly larger than 1 Gyr, the Ma05 models
predict $\mathrm{m_{0.55}}-\mathrm{m_{1.40}}$ NIR-optical colours that
are on average 0.5 mag redder than observed, for the observed optical
$\mathrm{m_{0.55}}-\mathrm{m_{0.70}}$ colours.\footnote{Similar
  conclusion is reported by \cite{conroy_gunn_10} concerning the
  colours of their sample of PSB galaxies.}

We note that the inclusion of dust with a standard effective
attenuation curve in the models does not alleviate the tension between
the Ma05 models and observations. We illustrate this in
Fig. \ref{fig:diag_optNIR_SSP} by drawing arrows which represent the
offset to apply to each model when an optical extinction of
$A_V=1$~mag and an effective attenuation law
$\tau_\lambda\propto\lambda^{-0.7}$ \citep[following][for an
SSP]{charlot_fall00} are applied. As expected, the effect of dust is
to move model tracks to redder colours: this would worsen the
disagreement in the index-colour planes. The reddening vector is
approximately parallel to the model tracks in the colour-colour plane:
accounting for dust attenuation in the models would not substantially
change the (dis)agreement with the observations in terms of colours,
although it would require even lower metallicities and ages.

\begin{figure}
\centerline{
\includegraphics[width=\columnwidth]{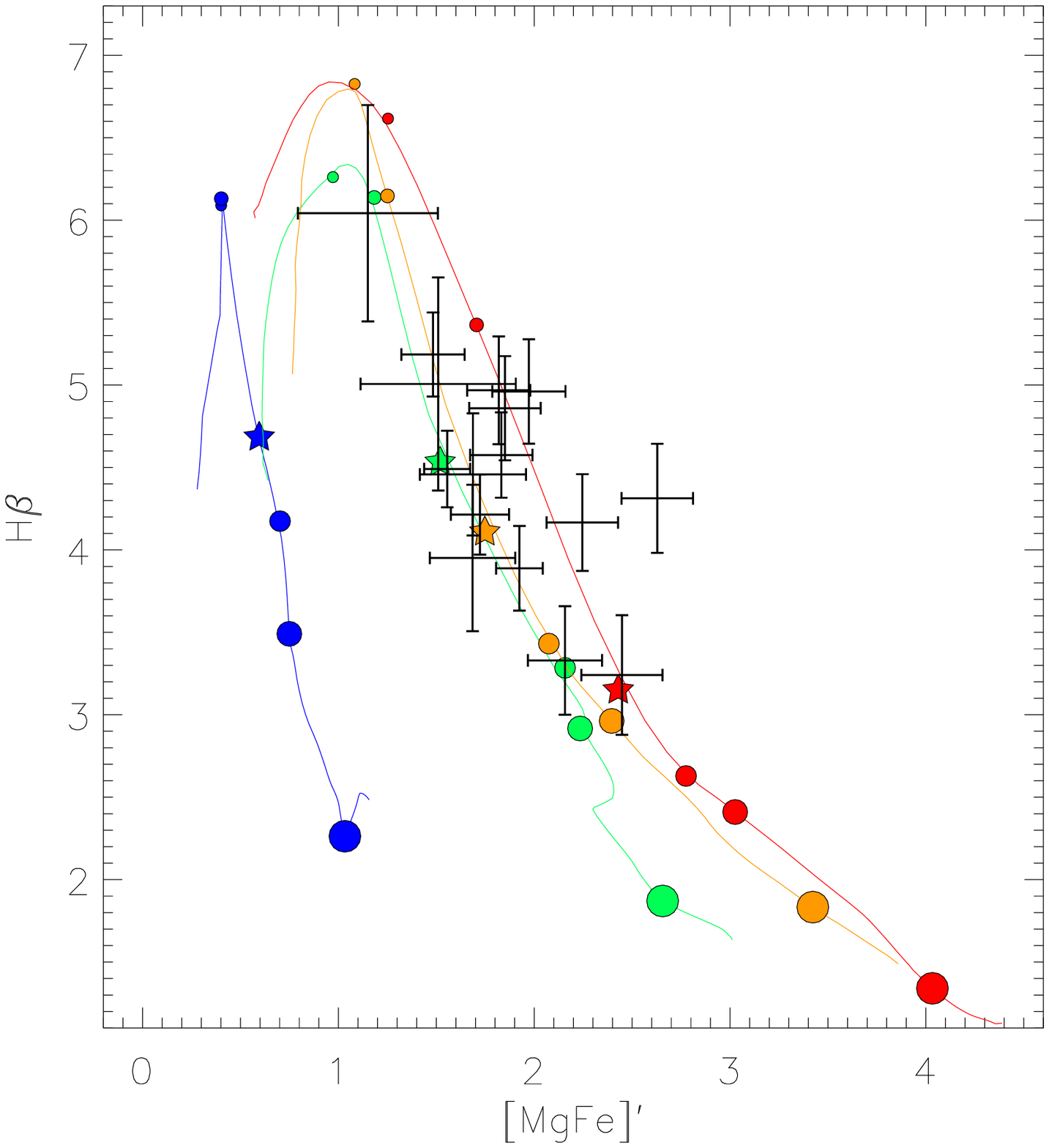}
}
\caption{H$\beta$ vs [MgFe]$^\prime$ diagnostic plot, showing that our
  galaxies need to have metallicity not lower than $\approx$ 40\%
  solar in order to be consistent with the observed strength of metal
  sensitive absorption lines. The four evolutionary tracks in colour
  are from BC03 models, colour coded and marked as in
  Fig. \ref{fig:diag_optNIR_SSP}.}\label{fig:Hbeta_MgFep}
\end{figure}
The hypothesis of a metallicity as low as 1/20 solar is not only
unlikely to apply to a full sample of relatively massive galaxies with
$M^*\approx 5\times10^{10} \mathrm{M_\odot}$, but is also inconsistent
with direct metallicity constraints which can be derived from
absorption indices. In Table \ref{tab:sample} we already showed
metallicity estimates derived from the vast library (based on BC03)
and bayesian method of \cite{gallazzi+05}: these estimates range from
0.2 $\mathrm{Z_\odot}$ minimum to 2.3 $\mathrm{Z_\odot}$ maximum,
roughly uniformly distributed. In Fig. \ref{fig:Hbeta_MgFep} we show
that metallicities as low as 1/20 solar are indeed ruled out already
from a simple comparison of standard absorption line indices on which
different models are in substantial agreement, namely [MgFe]$^\prime$
and H$\beta$. [MgFe]$^\prime$ in particular is chosen as optimal
metallicity indicator following \cite{thomas+03} who demonstrated its
insensitivity to $\alpha/$Fe abundance ratio; H$\beta$ serves to
constrain the age of the stellar population and hence alleviate the
well known age-metallicity degeneracy. Fig. \ref{fig:Hbeta_MgFep}
shows the indices measured in our galaxies as crosses with error bars,
while tracks of different colours are derived from BC03 SSPs in the
age range 50 Myr -- 13.6 Gyr (symbols are the same as in
Fig. \ref{fig:diag_optNIR_SSP}). Four different metallicities, 1/50,
0.4, 1 and 2.5 solar are plotted in purple, blue, green orange and
red, respectively. It is apparent from this plot that metallicities
significantly lower than 0.4 solar are inconsistent with the
data-points. We have checked that the same conclusions are obtained if
indices based on BC03 models are replaced with those computed by
Thomas and collaborators \citep{thomas+03,thomas+04,korn+05}: the only
difference between the two sets of models is that Thomas' models shift
the lower limit allowed by the data to slightly higher metallicities.

\subsection{Effects of composite stellar populations}\label{sec:CSPcomp}

A possible explanation why the strong TP-AGB features predicted by
Ma05 models are not seen in our spectra is that in the NIR there is a
substantial contribution by old stars, which are almost completely
outshined by the younger population in the optical bands. In fact, it
is quite unlikely that the entire stellar population of galaxies as
massive as several $10^{10}\mathrm{M_\odot}$ is formed in a burst
$\approx 1$~Gyr old and a more realistic scenario is that the burst
shines atop an old component. To explore this hypothesis we have
repeated the test presented in Fig. \ref{fig:diag_optNIR_SSP} using
composite stellar populations (CSP) instead of SSPs.

\begin{figure*}
\centerline{
\includegraphics[width=\textwidth]{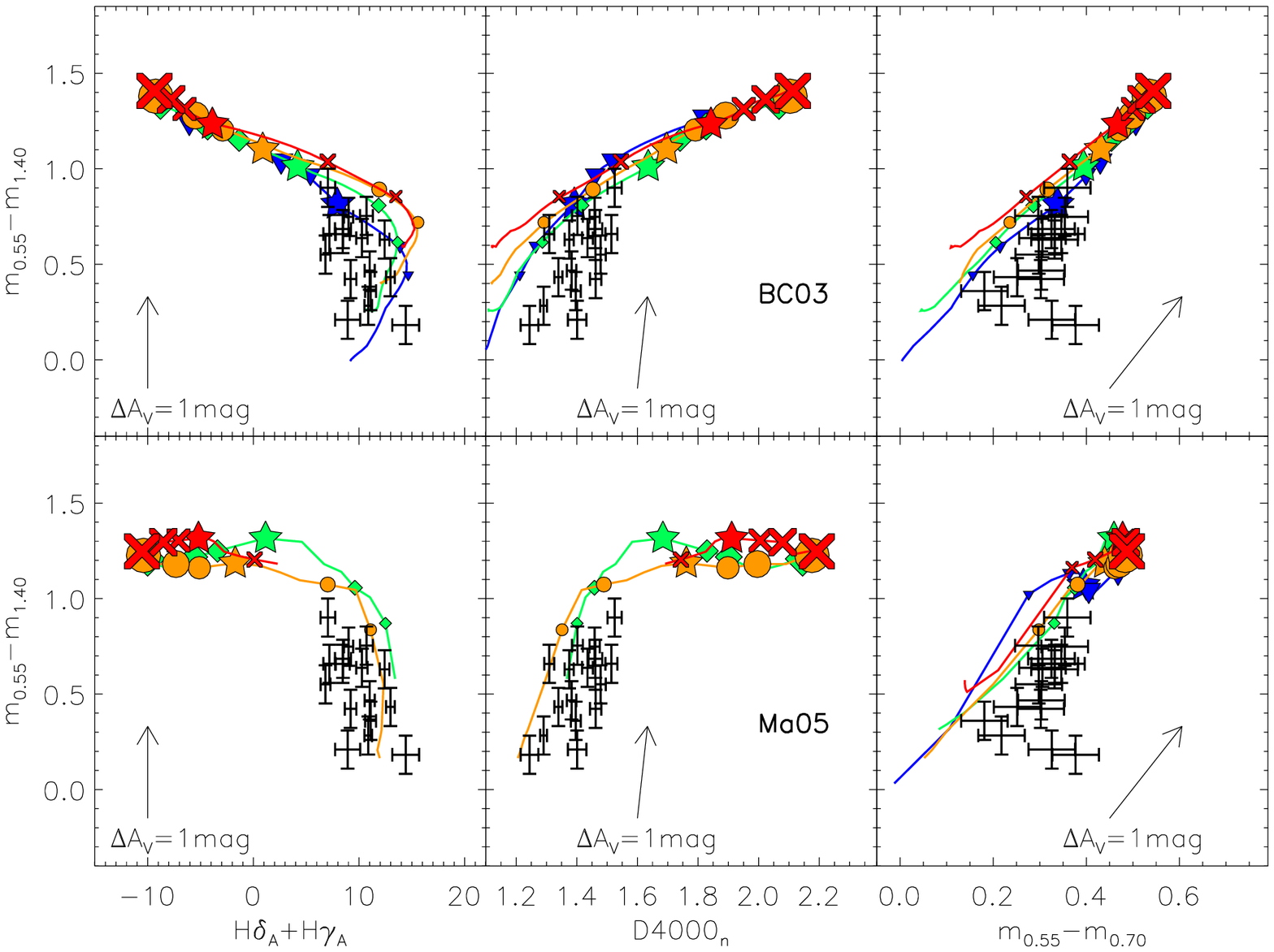}
}
\caption{Same as Figure \ref{fig:diag_optNIR_SSP}, but now using the
  composite SPs that are referred to as CSP-S in the text: a 10 Gyr
  old, solar metallicity component is combined with SSPs with the same
  properties as (and represented as) in Figure
  \ref{fig:diag_optNIR_SSP}. The ratio of masses \emph{as formed} in
  the fixed ``old'' vs variable ``young'' components is 9:1. For both
  BC03 and Ma05 models, such a combination can reproduce the spectral
  indices with ``young'' components younger than $\approx 0.5-1$ Gyr
  (even more so for the Ma05 models); however the optical-NIR colours
  cannot be reproduced by these models. Inclusion of standard dust
  attenuation (illustrated by the arrows) does not help also in this
  case. }\label{fig:diag_optNIR_cSP}
\end{figure*}
In a first experiment, each CSP is composed of two components: a fixed
10-Gyr old, solar metallicity SSP (which we will call ``old
component'' in the following) and an SSP of variable age, from 50 Myr
to 13.6 Gyr, and metallicity, from 1/20 to 2--2.5 solar (which we will
call ``young'' component or ``burst'', and corresponds to the SSPs of
Figure \ref{fig:diag_optNIR_SSP}). In the following we refer to these
models as CSP-S, where ``S'' indicates that the metallicity of the old
component is solar. This choice appears reasonable considering that
$Z\gtrsim\mathrm{Z_\odot}$ is typical for the old component of massive
galaxies with $M^*\gtrsim 10^{10}\mathrm{M_\odot}$
\citep[e.g.][]{gallazzi+05,thomas+05}. We have varied the relative
fraction of (initial) mass in the old component and in the burst from
0:1 to 0.9:0.1. Fig. \ref{fig:diag_optNIR_cSP} shows the same
diagnostic plots we already used to analyse pure SSPs
(Fig. \ref{fig:diag_optNIR_SSP}), but now plotting CSP-S models for
the extreme ``contamination'', i.e. for 90\% of the formed mass in the
old component and 10\% in the burst \citep[as was also assumed
by][]{conroy_gunn_10}. Symbols and colours are the same as in
Fig. \ref{fig:diag_optNIR_SSP} and refer to age and metallicity of the
variable burst component alone. CSP-S models, either BC03 or Ma05, can
easily reproduce the distribution of data-points in
$\mathrm{H\delta_A+H\gamma_A}$ and $\mathrm{D4000_n}$: the
contamination by the old component is compensated by a younger burst
component ($\lesssim 500$~Myr) with respect to those that provided the
best match in the pure burst (SSP) models. Such a combination would
easily explain why we do not see the sharp NIR features due to TP-AGB
stars predicted by Ma05: not only part of the NIR emission is provided
by the old and TP-AGB free stellar population, but also the young
component is younger that the peak of the TP-AGB phase.

\begin{figure*}
\centerline{
\includegraphics[width=\textwidth]{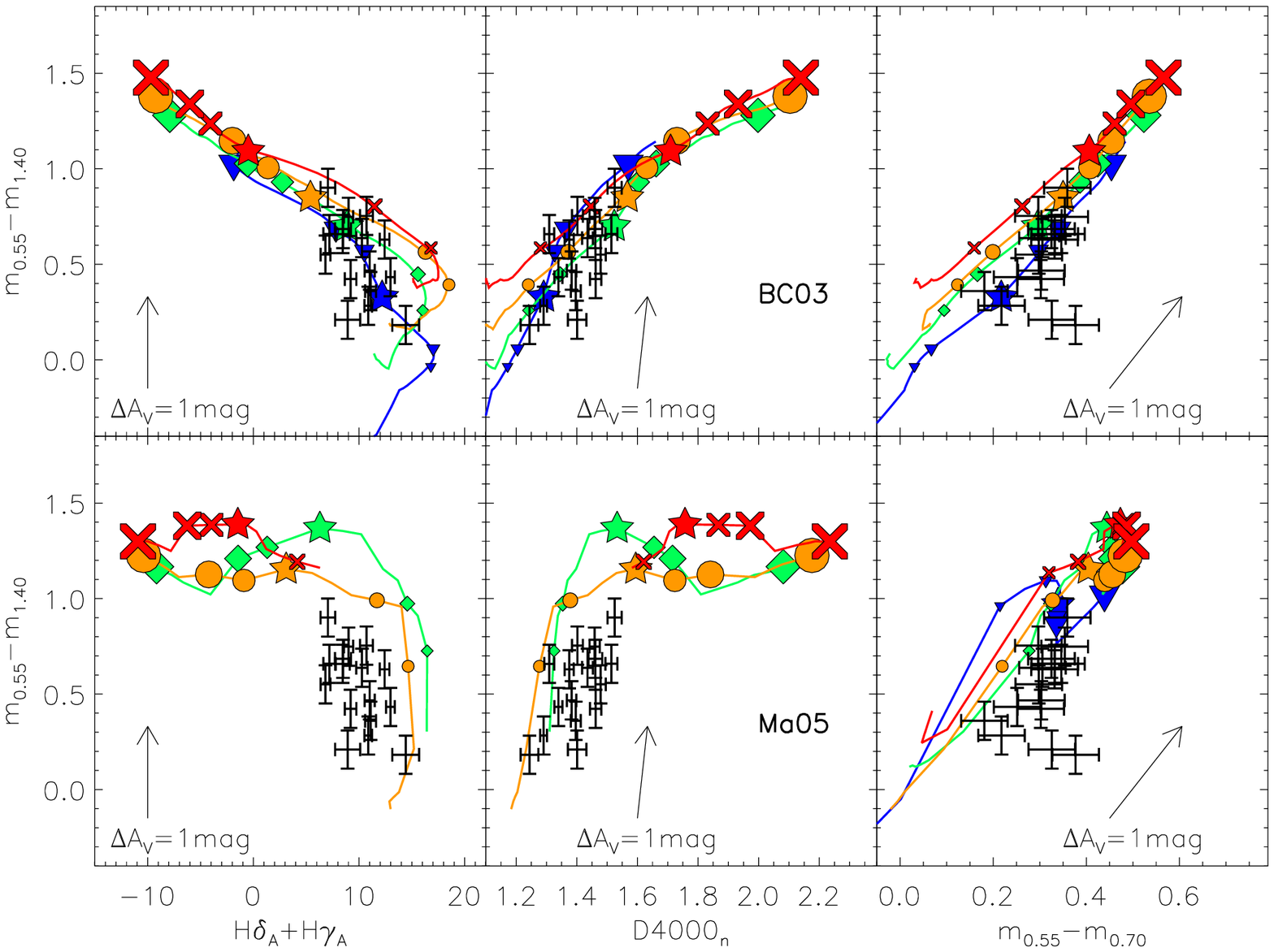}
}
\caption{Same as Figure \ref{fig:diag_optNIR_cSP}, but now using CSP-S
  with a ratio of masses \emph{as formed} in the fixed ``old'' vs
  variable ``young'' components of 2:1. This figure shows that a
  substantial underlying old stellar population is compatible with
  most of the observations if BC03 models are used, but produces
  inconsistent colours if the Ma05 are adopted
  instead.}\label{fig:diag_optNIR_cSP_20}
\end{figure*}

However, problems arise when optical-NIR colours are considered. The
addition of an old component at solar metallicity to a young burst
yields significantly redder optical-NIR colours with respect to a pure
burst. As a consequence, BC03 models, which provided a decent match in
the case of a pure SSP, are now shifted to inconsistently red colours
with respect to the observations. In the case of Ma05 models, which
already had troubles being too red in optical-NIR colours at almost
any age, they become inconsistent with the observed colours of almost
all galaxies. Mixtures with an intermediate mass ratio between the old
population and the burst produce also intermediate results in terms of
indices and colours between the pure burst models
(Fig. \ref{fig:diag_optNIR_SSP}) and the CSP-S ones presented in
Fig. \ref{fig:diag_optNIR_cSP}. In fact, as shown in
Fig. \ref{fig:diag_optNIR_cSP_20}, a more moderate ratio between
``old'' and ``burst'' stars of 2:1 yields a decent match between the
BC03 models and the observations. As for the Ma05 models, however, the
fact that already SSPs are too red in optical-NIR colours results in
no combination of ``old'' vs ``burst'' stars being able to reproduce
the observed colours.

\begin{figure*}
\centerline{
\includegraphics[width=\textwidth]{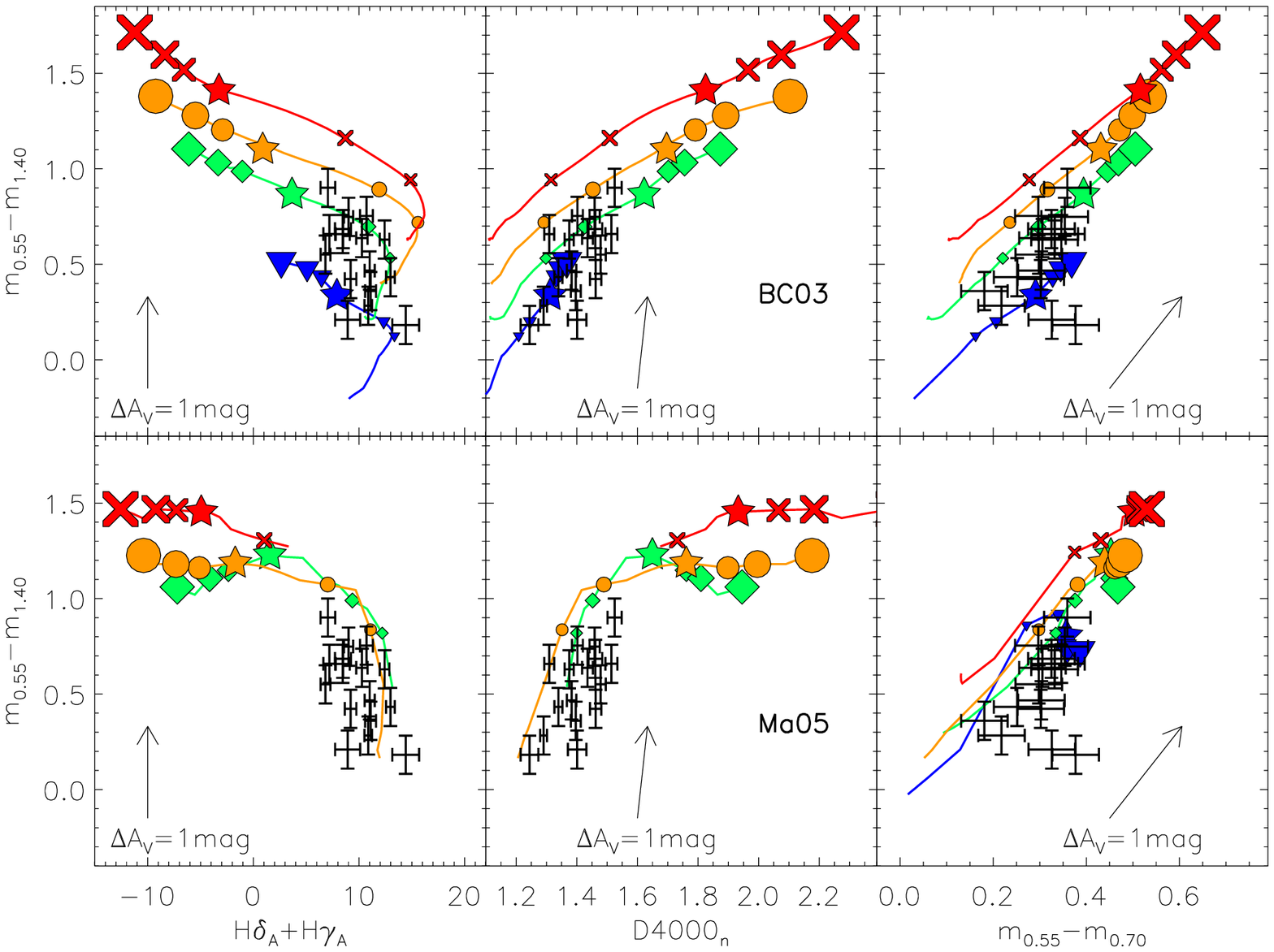}
}
\caption{Same as Figure \ref{fig:diag_optNIR_cSP} for CSP-Y models
  (see text), composite SPs with the ``old'' component having the same
  metallicity as the ``young'' one, according to the corresponding
  track.}\label{fig:diag_optNIR_cSP_sameZ}
\end{figure*}

We repeated the experiment with the 9:1 ratio, but changing the
metallicity of the old component from $Z=\mathrm{Z_\odot}$ fixed to a
variable metallicity that matches the one of the ``young''
component. We tag these models as CSP-Y, where ``Y'' indicates that
the metallicity of the old component is the same as in the young
component. The results are shown in
Fig. \ref{fig:diag_optNIR_cSP_sameZ}. BC03 models
(Fig. \ref{fig:diag_optNIR_cSP_sameZ} \textit{top}) are able to cover
most of the parameter space occupied by the data-points; however a
significant number of galaxies are too blue in
$\mathrm{m_{0.55}}-\mathrm{m_{1.40}}$ with respect to models at given
$\mathrm{D4000_n}$. Moreover, the metallicities implied by colours are
systematically lower (and to large extent inconsistent) with those
implied by the absorption indices alone. As one can argue in the light
of the results of pure SSP and 2:1 ratio models, less extreme
fractions of the old component can provide a much better match to the
data.  As for the Ma05 models, Fig. \ref{fig:diag_optNIR_cSP_sameZ}
\emph{bottom} shows that not even with this kind of mix it is possible
to match the distribution of $\mathrm{m_{0.55}}-\mathrm{m_{1.40}}$ vs
$\mathrm{D4000_n}$, the optical-NIR colours of the models being too
red with respect to the data. We reiterate that any dust with the most
commonly adopted attenuation curves would only worsen this
disagreement.

In summary, adding a substantial fraction of old stellar populations
to reduce the impact of TP-AGB stars predicted by Ma05 in the NIR may
indeed explain the lack of observed NIR features, while remaining
consistent with the observed optical spectral indices. However, such a
combination fails to reproduce the observed optical-NIR colours at
given $\mathrm{H\delta_A+H\gamma_A}$, $\mathrm{D4000_n}$ and optical
colours, no matter if the metallicity of the ``old'' component is the
same as the one of the ``young'' component or is fixed at the solar
value (as it is probably more realistic for an old stellar population
in a system of a few times $10^{10} \mathrm{M_\odot}$). We cannot
exclude that more complicated and \emph{ad hoc} star formation
histories and metallicity mixtures might be able to produce a closer
match between observations and Ma05, but it appears unlikely that such
a fine tuning actually applies to all 16 galaxies in the sample. On
the other hand, BC03 models appear to reproduce the observations much
more easily by assuming SSPs or CSPs with even massive ``old''
components of variable sub-solar metallicity; in order for models with
an ``old'' component at solar metallicity to match the observations,
``old'' fractions less than 0.9 are required in most cases: it is
sufficient to lower such fraction to 0.6-0.7 in order to obtain a
generally good agreement.

\begin{figure*}
\centerline{
\includegraphics[width=\textwidth]{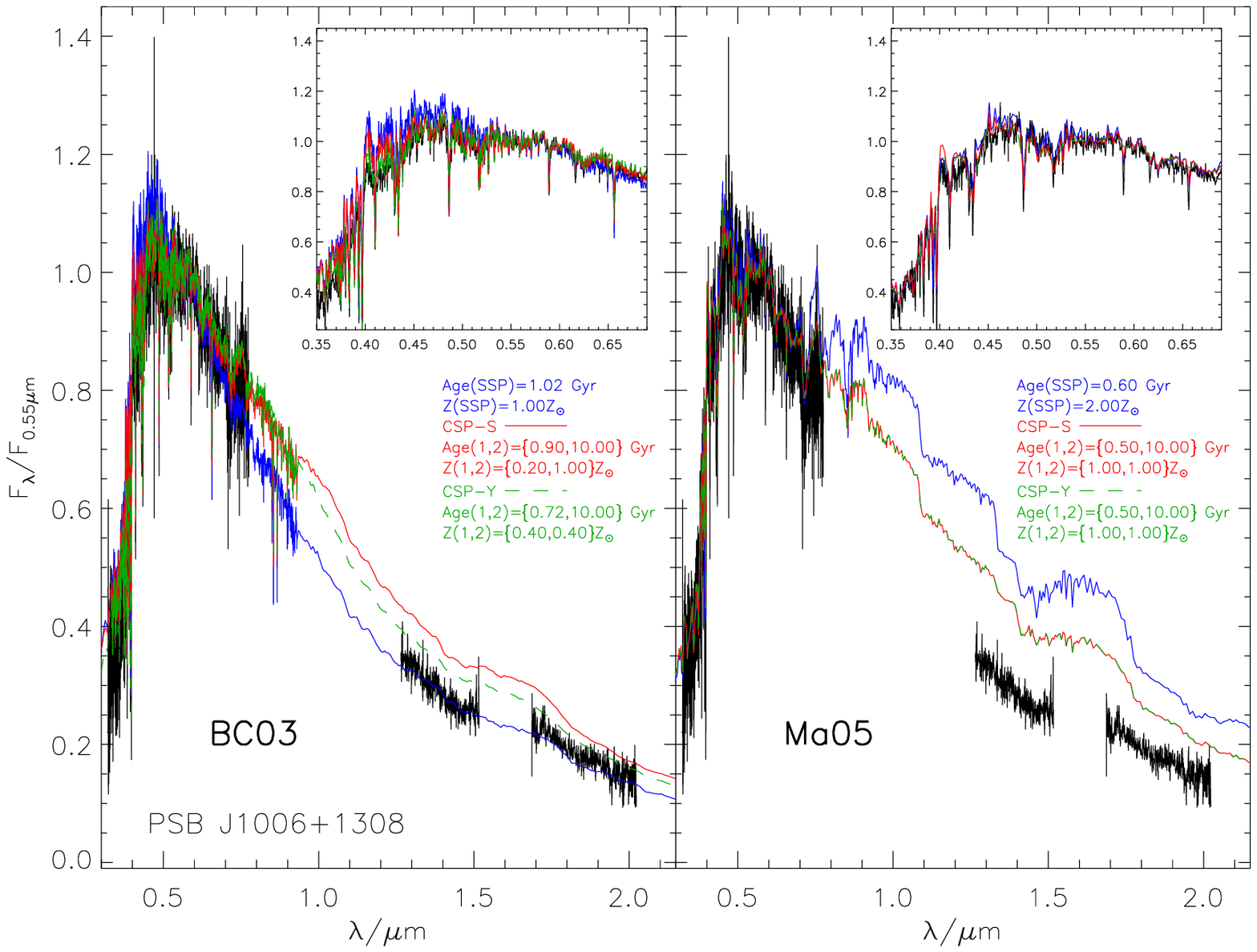}
}
\caption{Example of how different models perform in reproducing the
  SED of a PSB galaxy, specifically PSB\,J1006+1308, whose
  \textit{observed} spectrum is shown in black. In the left-hand panel
  models based on BC03 are overplotted, while models based on Ma05 are
  shown in the right-hand panel: the blue lines are used for pure SSP,
  red solid lines for CPS-S (90\% 10-Gyr old component at solar
  metallicity) and green dashed lines for CSP-Y (90\% 10-Gyr old
  component at the same metallicity as the young component). The
  models plotted here are chosen from each of the simple grids
  presented in the text and in Fig. \ref{fig:diag_optNIR_SSP},
  \ref{fig:diag_optNIR_cSP} and \ref{fig:diag_optNIR_cSP_sameZ} as
  those that best fit the set of five optical stellar absorption
  features sensitive to age and metallicity defined in Gallazzi et
  al. (2005). All models can reproduce the spectral absorption
  features remarkably well (see insets) and agree on the presence of a
  young component approximately at the peak of the TP-AGB
  contribution. However their extrapolations to the NIR regime differ
  significantly. Most notably, while the inclusion of a substantial
  contribution from an old component can help reduce the strong NIR
  features predicted by Ma05 models, the NIR-to-optical ratio
  predicted by these models remains nevertheless too
  high.}\label{fig:example_cSP_J1006}
\end{figure*}

We further illustrate these points by directly comparing SSP and CSP
model spectra with an actual spectrum of our PSB sample in
Fig. \ref{fig:example_cSP_J1006}, where PSB\,J1006$+$1308 is used as a
mere example which optimally serves to this goal, as it is broadly
representative of the full sample. For both BC03 (shown in the
left-hand panel) and Ma05 (in the right-hand panel), we pick the best
fitting models from each of the three grids of SSP, CSP-S and CSP-Y
(assuming old:young=9:1). They are overplotted to the observed
spectrum (solid black line), in blue for the SSP, in solid red for the
CSP-S and dashed green for the CSP-Y. The ``best fitting'' models are
those yielding the minimum $\chi^2$ over the optimal set of optical
absorption indices defined in \cite{gallazzi+05} ($\mathrm{D4000_n}$,
$\mathrm{H\delta_A+H\gamma_A}$, $\mathrm{H\beta}$,
$\mathrm{[MgFe]^\prime}$, $\mathrm{[Mg_2Fe]}$). Due to the coarse grid
in metallicity, the use of fixed old fractions and the neglect of dust
attenuation, we warn the reader not to over-interpret these fits in
terms of derived physical parameters. Nonetheless, the following key
observations can be derived from
Fig. \ref{fig:example_cSP_J1006}. \emph{i)} All best fit models can
reproduce the observed optical features extremely well (see
insets). The slope of the optical continuum is also generally well
reproduced, although for BC03 the SSP fit is too blue in the optical
range, which might be due either to a small mismatch in metallicity
because of the coarse grid or to small amounts of dust, which are not
included in our simple models. \emph{ii)} The NIR spectrum of the best
fitting Ma05 SSP, with an age of 0.6 Gyr and metallicity 2 times
solar, has the typical NIR boosting due to the Ma05 implementation of
the TP-AGB phase, clearly recognisable from the sharp features around
1.41 and 1.77~$\mu$m. However, no evidence of such features appears in
the observed spectrum; moreover, the observed spectrum is almost a
factor of two dimmer than the Ma05 SSP in the NIR. In contrast, the
BC03 SSP provides a very good fit in the NIR both in terms of (lack
of) features and of overall flux. \emph{iii)} The use of CSPs with a
substantial fraction (90\%) of 10-Gyr old stars yields a relative
enhancement in the NIR over optical for BC03 models, for both CSP-S
and CSP-Y. The plot suggests that a smaller old fraction and lower
metallicity than the best fitting pure SSP might provide a very close
match to the NIR-to-optical ratio. CSPs for Ma05 (which by coincidence
converge to the same solution for both CSP-S and CSP-Y, i.e. the solid
red line and the dashed green one overlap) do actually yield much
weaker NIR features and also lower NIR flux relative to
optical. However, this stays significantly higher ($\approx 50$\%)
than observed.

\subsection{The anomalous galaxies PSB\,J1150$-$0107 and
  PSB\,J1206$+$0918}\label{subsec:deviant_objs}

PSB\,J1150$-$0107 and PSB\,J1206$+$0918 are characterised by extremely
blue optical-NIR colours ($\mathrm{m_{0.55}}-\mathrm{m_{1.40}}\lesssim
0.2$) and intermediate/red optical colours
($\mathrm{m_{0.55}}-\mathrm{m_{0.70}}\gtrsim 0.3$). When plotted on
the colour-colour diagrams of Fig. \ref{fig:diag_optNIR_SSP},
\ref{fig:diag_optNIR_cSP}, \ref{fig:diag_optNIR_cSP_20} and
\ref{fig:diag_optNIR_cSP_sameZ}, they significantly depart from the
sequences both of observed data-points and model tracks. None of them
is able to reproduce such a combination of intermediate/red optical
colours and very blue optical-NIR ones: for the given
$\mathrm{m_{0.55}}-\mathrm{m_{1.40}}$, an offset of $-0.1$ to $-0.2$
mag in $\mathrm{m_{0.55}}-\mathrm{m_{0.70}}$ would be required to
bring them back in agreement with the models, or, for the given
$\mathrm{m_{0.55}}-\mathrm{m_{0.70}}$, an offset of $\gtrsim 0.3$ mag
in $\mathrm{m_{0.55}}-\mathrm{m_{1.40}}$. From the arrows plotted in
the colour-colour diagrams, it is immediate to see that a standard
effective attenuation \`a la \cite{charlot_fall00} is unable to move
the models closer to these two objects: in fact, one would need an
effective attenuation curve which is extremely flat, thus causing
similar attenuation at 0.70 and 1.40~$\mu$m. With standard dust
grains, for which the absorption cross section decreases with
wavelength, this can be only obtained by invoking a peculiar geometry
in which back scattering plays a major role.

Significant residual star formation is witnessed in PSB\,J1150$-$0107
by EW(H$\alpha$)$=-2.43\pm0.24$ and a marginal detection at 22~$\mu$m
in the WISE all-sky survey. PSB\,J1206$+$0918, however, does not
display any sign of star formation, with EW(H$\alpha$)$=-0.65\pm0.26$
and no detection at 22~$\mu$m. This suggests that residual star
formation or processes related to nuclear activity are not the cause
of the anomaly common to these two galaxies.

We note that the shape of the spectra of these two galaxies is
peculiar in two main aspects: \textit{i)} the continuum appears to
bend down, i.e. the slope becomes steeper, going from $\approx 0.55$
to 0.8~$\mu$m, whereas all other galaxies display an approximate
constant slope or even an up-bending shape in this region;
\textit{ii)} a significant and featureless drop long-ward of $\approx
0.5$~$\mu$m~is observed instead of the Fe and Mg absorption complexes,
which is probably due to the low metallicity of these two galaxies,
among the lowest of the entire sample. These two features appear to be
responsible for the anomalous colours, by making the spectral slope
between 0.7 and 1.4~$\mu$m too steep relative to the one in the
optical range 0.55--0.70~$\mu$m. The fact that models fail to
reproduce these colours may point to problems in the low metallicity
regime, below $\approx 0.5 \mathrm{Z_\odot}$.

We note that, according to the Ma05 models, the largest enhancement
and the most dramatic features in the NIR are expected for
$Z\lesssim\mathrm{Z_\odot}$. Instead these two low-metallicity
galaxies exhibit bluer optical-NIR colours than even "TP-AGB light"
models (e.g., BC03).

\subsection{Comparison with Kriek\etal (2010)}

As already mentioned in Sec. \ref{sec:intro}, \cite{kriek+10} have
obtained a detailed (rest-frame) UV-optical-NIR SED for a sample of 62
PSB galaxies observed in the \textit{NEWFIRM} Medium Band Survey
(NMBS) in the redshift range 0.7--2. Their composite SED is
essentially a stack of the SED of their sample. They applied maximum
likelihood fitting using both BC03 and Ma05 on different wavelength
ranges (optical only or optical plus NIR) and, similarly to the
results we have presented in this work, they found that Ma05 cannot
reproduce the blue optical-NIR colours observed, whereas BC03 provide
an almost perfect match.
\begin{figure}
\includegraphics[width=\columnwidth]{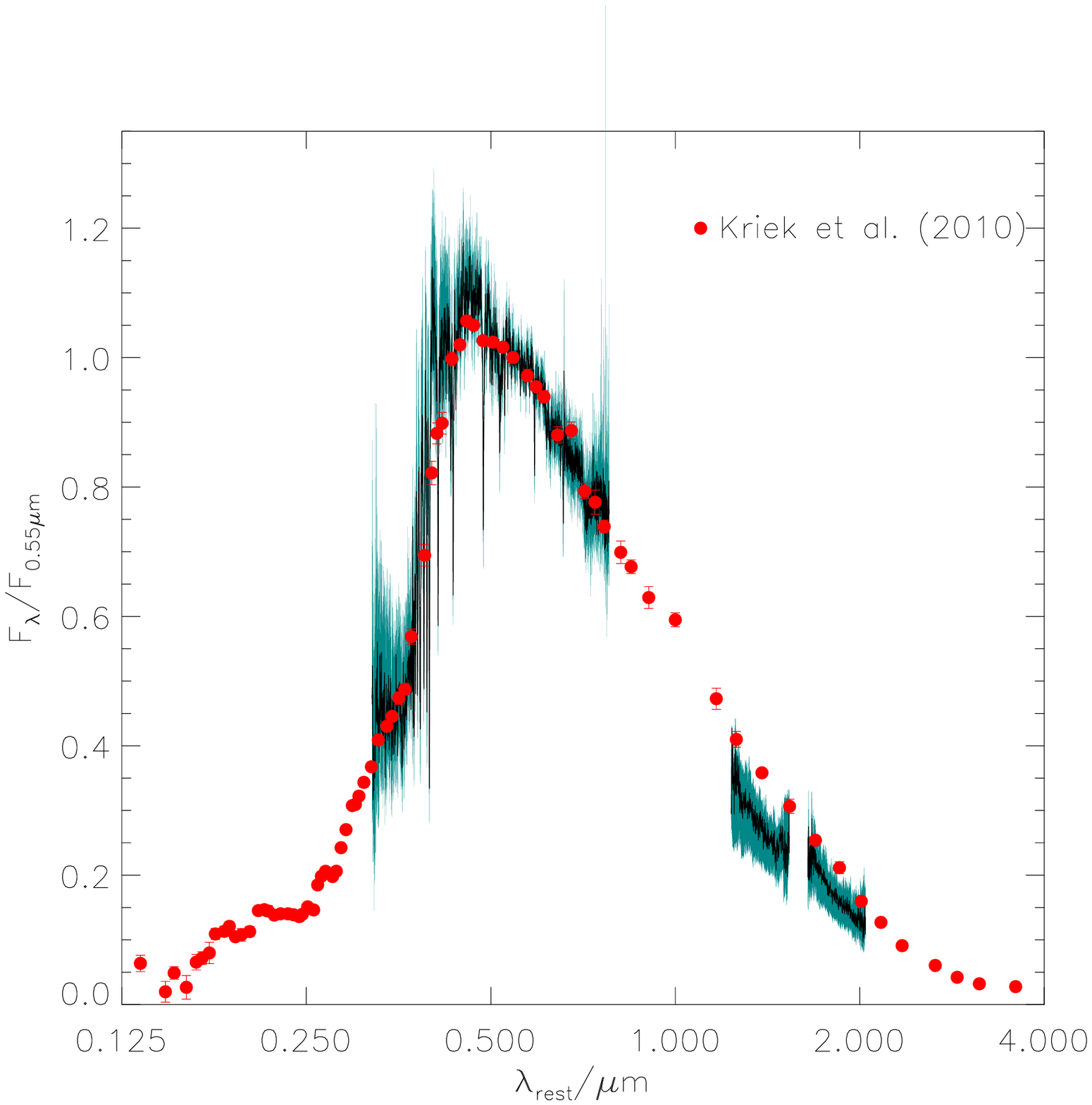}
\caption{Comparison between the stacked SED from our spectra and from
  the post-starburst sample from the NMBS of Kriek\etal (2010), shown
  as red points with error bars. The black line is the median spectrum
  of our 16 galaxies, whose spectra had been previously normalised at
  5500~\AA. The shaded area shows the 16--84 percentile range of the
  data. The two samples yield consistent results (although our sample
  appears marginally biased towards bluer SEDs relative to
  Kriek\etal~2010) and, most importantly, agree on the fact that there
  is no need for boosted NIR emission by TP-AGB as implied by
  Ma05.}\label{fig:comp_Kriek}
\end{figure}

In Fig.  \ref{fig:comp_Kriek} we compare the SED published by
\cite{kriek+10}, represented by the red points with error bars, with
the stack of our 16 spectra. Both SEDs are normalised at 5500~\AA. The
black line is the median of our 16 spectra at each wavelength, while
the shaded area displays the 16--84 inter-percentile range. The
similarity of the stacked SEDs from the two works is striking,
typically consistent within 1--1.5 $\sigma$.  Our spectra are slightly
bluer and steeper in the range 5000~\AA~to $2~\mu$m, which might
derive from some relative bias in the respective sample selections.,
with our objects being somehow more extreme (e.g. in terms of burst
fraction, see also Sec. \ref{sec:summary}).  In fact, \cite{kriek+10}
define their sample based on k-corrected rest-frame colour cuts which
are hardly comparable to our spectroscopic criteria.  Alternatively,
this comparison might indicate that NIR-bright evolved (TP-AGB) stars
have a slightly larger influence on the SED of PSB galaxies at
$z\gtrsim 0.7$ than at $z\approx 0.2$, consistently with expectations,
e.g., of \cite{melbourne+12}. Apart from this minor colour difference,
our detailed spectroscopic study of individual PSB galaxies confirms
and reinforces the previous results by \cite{kriek+10}, based on
stacking and medium band photometry and at a very different age of the
Universe.

\section{Summary and concluding remarks}\label{sec:summary}

We have presented the first spectroscopic study of post-starburst
galaxies covering both the optical (SDSS) and NIR (our new VLT-ISAAC
observations) wavelength ranges, which was specifically designed to
detect the two main signatures of the presence of TP-AGB stars
predicted by the \cite{maraston05} models in contrast to ``TP-AGB
light'' models (such as BC03): \textit{i)} the strong NIR absorption
features caused by carbon composite molecular band-heads; and
\textit{ii)} the boosting of the NIR flux. To this goal, we have
carefully selected a sample of 16 galaxies with post-burst ages
corresponding to the peak contribution by TP-AGB stars to the NIR flux
(0.5--1.5 Gyr), in a broad range of stellar metallicities ($\approx
0.2$--2.5 $\mathrm{Z_\odot}$). These galaxies are at high enough
redshift (0.15--0.25) to allow clean measurements of the NIR features
at rest wavelengths 1.41 and 1.77~$\mu$m through the atmospheric H and
K windows. Despite these ideal conditions for the detection of the
characteristic signatures of TP-AGB stars predicted by the Ma05
models, the 16 galaxies in our sample all display smooth, featureless
NIR spectra and blue optical-NIR colours. Interestingly, these
observed spectral properties are generally well reproduced using the
BC03 models assuming simple (i.e. single-age) stellar populations.

We have investigated whether the observations could be reproduced with
Ma05 models as well, by combining stellar populations of different
ages.  We find that diluting the light of a young (i.e. a few 100 Myr
old) starburst with that of a massive old component allows the Ma05
models to reproduce the typical ``E+A'' optical spectra of
post-starburst galaxies characterised by strong Balmer absorption
lines and a well developed D4000 break. At the same time, such models
exhibit very weak NIR features, in agreement with the observations,
because TP-AGB stars from the young component have not developed yet,
while the NIR emission from old stellar populations is
smoother. However, these models still cannot account for the data,
because the predicted ratio of NIR to optical flux is significantly
larger than observed at the metallicity of our galaxies. Hence, the
observed lack of deep NIR absorption features and the modest ratio of
NIR to optical flux in our data cannot be simultaneously accounted for
by a simple ``light dilution'' effect.  We note that the observation
of metal absorption lines at optical wavelength does not support the
idea that the faint NIR continuum (and the consequent dilution and/or
absence of strong NIR absorption features) could arise from extremely
low metallicities.

The Ma05 models predict too red optical-NIR colours for any choice of
ratio between the old and the young component. On the contrary, models
based on BC03 fail for ``old'' mass fractions as high as 90\%, but
produce colours that are consistent with (most of) the observations
for ``old'' mass fractions up to 60--70\%, which leaves enough space
to accommodate more reasonable SFHs than a (possibly unrealistic)
single burst. We note that the work of \cite{pracy+12} indicates old
fractions close to 90\% for their seven PSB galaxies, which would be
inconsistent with our SEDs. However, this estimate might be biased by
their assumption of fixed solar metallicity. Moreover, and probably
more importantly, it is possible that our objects are somehow more
extreme in terms of burst fraction, because, due to our redshift
constraints, we select objects close to the faintest limit of the
spectroscopic sample of the SDSS. In this way we would select the
intrinsically most luminous PSB galaxies at $z\approx 0.2$, hence
those with the largest burst fraction (i.e. with the lowest ``old''
fraction): in fact, for a given total mass, a higher burst fraction
yields higher luminosity. The lack of strong colour gradients in most
of our PSB galaxies lends support to this hypothesis, suggesting that
the starburst substantially affected the full extent of the galaxy,
rather than the centre only. Finally, \cite{pracy+12} point out that
their sample is less extreme than typical samples of PSB galaxies in
the literature due to the small physical size of the spectroscopic
fiber (few hundred parsecs), which bias their selection towards
central rather than extended starbursts. In conclusion, the
consistency of our SEDs with BC03-based CSPs and ``old'' mass
fractions up to 60--70\% does not raise any problem for the physical
interpretation of our PSB galaxies using the BC03 models.

We also find that attenuation by interstellar dust, using standard
prescriptions, can only worsen the disagreement (when present) between
spectral evolution models and the observed optical-NIR colours of
post-starburst galaxies in our sample. This is particularly true for
the Ma05 models, which predict systematically redder colours than
observed. We note that heavier-than-implemented attenuation by {\em
  circumstellar} dust of the most luminous C stars about to eject
their envelopes could potentially reduce the contribution by these
stars to the integrated spectrum and perhaps improve the agreement of
the Ma05 models with observations.\footnote{The Ma05 models rely on
  period-averaged spectra of TP-AGB stars assembled by
  \cite{lancon_mouhcine_02}, who recommend reddening these spectra
  before use in population synthesis models (see their section
  3.3). Whether this or a different attenuation has been included
  cannot be asserted from the description of the TP-AGB spectra in
  section 3.3.2 of Ma05. In the BC03 models, the circumstellar
  attenuation of TP-AGB stars experiencing superwinds is accounted for
  in an empirical way (see section 2.2.4 of BC03 for detail).} A more
accurate theoretical and empirical assessment of the effects of
circumstellar dust absorption is definitely key to make progress in
the SED modelling of the stellar population affected by TP-AGB stars
\citep[e.g.][]{srinivasan+09,melbourne+12,meidt+12}.

The results presented in this work confirm the previous claim by
\cite{kriek+10} that BC03, and more generally ``TP-AGB light'' models,
provide a better agreement with the spectral properties of galaxies
hosting a stellar population at the peak of the TP-AGB phase than Ma05
models. The important novelty of our work is to base these conclusions
on a higher-resolution, high-quality spectrophotometric dataset and an
optimal spectroscopic sample selection.

Somehow in contrast with these studies and our work in particular,
\cite{lyubenova+12} have detected the 1.77~$\mu$m feature in a couple
of Magellanic Cloud's globular clusters and found them in agreement
with Ma05 predictions. This result calls for more observations in
order to obtain a representative sample of globular clusters over a
broader range of metallicities (the two in which the detection was
obtained are both at metallicity between 0.2 and 0.6 solar) and beat
down stochasticity effects: in fact, a single (TP-)AGB star can
produce a substantial fraction of the total cluster NIR luminosity in
the mass range explored. One further possibility to reduce the
apparent disagreement between the \cite{lyubenova+12} results and ours
might reside in a technical aspect of their challenging observations:
the addition of targeted AGB stars outside of the central region to
integrate the central spectrum and obtain full spatial coverage of the
globular clusters might lead to AGB stars being over-represented with
respect to the stars that contribute the bulk of the diffuse emission.

In the light of this complex and partly controversial observational
framework, the apparent agreement of the BC03 models (of either simple
or composite stellar populations) with the observations presented in
this paper should not be necessarily interpreted as an indication that
the description of the TP-AGB phase in these models is adequate.  New
models incorporating more recent developments in the theory and
observation of TP-AGB stars \citep[e.g.][]{marigo+08,
  srinivasan+09,girardi+10,melbourne+12} are being assembled (Charlot
\& Bruzual, in preparation).  By providing a better account of
observations, new models of this type will allow deeper insight into
the physical properties of galaxies from spectral interpretations,
especially at high redshift, where a large number of galaxies is going
through the evolutionary phase affected by TP-AGB stars. In this
perspective, the present work provides key and very stringent
constraints for the new generation of stellar population synthesis
models.

\section*{Acknowledgments}
We thank the anonymous referee for useful comments and suggestions
that significantly improved the clarity of the presentation.\\
The Dark Cosmology Centre is funded by the Danish National Research
Foundation.\\
We also thank Paolo Cassata, Harald Kuntschner and Mariya Lyubenova
for useful comments. S.Z. acknowledges the generous hospitality of the
Dark Cosmology Centre and stimulating discussions with Jens Hjorth and
Sune Toft. D.P. acknowledges the kind hospitality at the
Max-Planck-Institut f\"ur extraterrestrische Physik.


\begin{thebibliography}{}

\bibitem[\protect\citeauthoryear{{Abazajian}, {Adelman-McCarthy},
  {Ag{\"u}eros}, {Allam}, {Allende Prieto}, {An}, {Anderson}, {Anderson},
  {Annis}, {Bahcall} et al.}{{Abazajian} et~al.}{2009}]{SDSS_DR7}
{Abazajian} K.~N.,  {Adelman-McCarthy} J.~K.,  {Ag{\"u}eros} M.~A.,  {Allam}
  S.~S.,  {Allende Prieto} C.,  {An} D.,  {Anderson} K.~S.~J.,  {Anderson}
  S.~F.,  {Annis} J.,  {Bahcall} N.~A.,    et al. 2009, \apjs, 182, 543

\bibitem[\protect\citeauthoryear{{Aihara} {et~al.}}{{Aihara}
  {et~al.}}{2011}]{SDSS_DR8}
{Aihara} H.,  {et~al.} 2011, \apjs, 193, 29

\bibitem[\protect\citeauthoryear{{Bell}, {McIntosh}, {Katz} \&
  {Weinberg}}{{Bell} et~al.}{2003}]{bell+03}
{Bell} E.~F.,  {McIntosh} D.~H.,  {Katz} N.,    {Weinberg} M.~D.,  2003, \apjs,
  149, 289

\bibitem[\protect\citeauthoryear{{Bruzual}}{{Bruzual}}{2007}]{bruzual07}
{Bruzual} G.,  2007, in {Vallenari} A.,  {Tantalo} R.,  {Portinari} L.,
  {Moretti} A.,  eds, From Stars to Galaxies: Building the Pieces to Build Up
  the Universe Vol.~374 of Astronomical Society of the Pacific Conference
  Series, {Stellar Populations: High Spectral Resolution Libraries. Improved
  TP-AGB Treatment}.
pp 303--+

\bibitem[\protect\citeauthoryear{{Bruzual} \& {Charlot}}{{Bruzual} \&
  {Charlot}}{2003}]{bc03}
{Bruzual} G.,  {Charlot} S.,  2003, \mnras, 344, 1000

\bibitem[\protect\citeauthoryear{{Cardelli}, {Clayton} \& {Mathis}}{{Cardelli}
  et~al.}{1989}]{cardelli+89}
{Cardelli} J.~A.,  {Clayton} G.~C.,    {Mathis} J.~S.,  1989, \apj, 345, 245

\bibitem[\protect\citeauthoryear{{Charlot} \& {Bruzual}}{{Charlot} \&
  {Bruzual}}{1991}]{charlot_bruzual_91}
{Charlot} S.,  {Bruzual} A.~G.,  1991, \apj, 367, 126

\bibitem[\protect\citeauthoryear{{Charlot} \& {Fall}}{{Charlot} \&
  {Fall}}{2000}]{charlot_fall00}
{Charlot} S.,  {Fall} S.~M.,  2000, \apj, 539, 718

\bibitem[\protect\citeauthoryear{{Chisari} \& {Kelson}}{{Chisari} \&
  {Kelson}}{2012}]{chisari_kelson12}
{Chisari} N.~E.,  {Kelson} D.~D.,  2012, \apj, 753, 94

\bibitem[\protect\citeauthoryear{{Conroy} \& {Gunn}}{{Conroy} \&
  {Gunn}}{2010}]{conroy_gunn_10}
{Conroy} C.,  {Gunn} J.~E.,  2010, \apj, 712, 833

\bibitem[\protect\citeauthoryear{{Conroy}, {Gunn} \& {White}}{{Conroy}
  et~al.}{2009}]{conroy+09}
{Conroy} C.,  {Gunn} J.~E.,    {White} M.,  2009, \apj, 699, 486

\bibitem[\protect\citeauthoryear{{Fukugita}, {Ichikawa}, {Gunn}, {Doi},
  {Shimasaku} \& {Schneider}}{{Fukugita} et~al.}{1996}]{fukugita_etal96}
{Fukugita} M.,  {Ichikawa} T.,  {Gunn} J.~E.,  {Doi} M.,  {Shimasaku} K.,
  {Schneider} D.~P.,  1996, \aj, 111, 1748

\bibitem[\protect\citeauthoryear{{Gallazzi}, {Charlot}, {Brinchmann}, {White}
  \& {Tremonti}}{{Gallazzi} et~al.}{2005}]{gallazzi+05}
{Gallazzi} A.,  {Charlot} S.,  {Brinchmann} J.,  {White} S.~D.~M.,
  {Tremonti} C.~A.,  2005, \mnras, 362, 41

\bibitem[\protect\citeauthoryear{{Girardi}, {Williams}, {Gilbert},
  {Rosenfield}, {Dalcanton}, {Marigo}, {Boyer}, {Dolphin}, {Weisz},
  {Melbourne}, {Olsen}, {Seth} \& {Skillman}}{{Girardi}
  et~al.}{2010}]{girardi+10}
{Girardi} L.,  {Williams} B.~F.,  {Gilbert} K.~M.,  {Rosenfield} P.,
  {Dalcanton} J.~J.,  {Marigo} P.,  {Boyer} M.~L.,  {Dolphin} A.,  {Weisz}
  D.~R.,  {Melbourne} J.,  {Olsen} K.~A.~G.,  {Seth} A.~C.,    {Skillman} E.,
  2010, \apj, 724, 1030

\bibitem[\protect\citeauthoryear{{Goto}}{{Goto}}{2005}]{goto_05}
{Goto} T.,  2005, \mnras, 357, 937

\bibitem[\protect\citeauthoryear{{Groenewegen}, {Sloan}, {Soszy{\'n}ski} \&
  {Petersen}}{{Groenewegen} et~al.}{2009}]{groenewegen+09}
{Groenewegen} M.~A.~T.,  {Sloan} G.~C.,  {Soszy{\'n}ski} I.,    {Petersen}
  E.~A.,  2009, \aap, 506, 1277

\bibitem[\protect\citeauthoryear{{Ilbert} {et~al.}}{{Ilbert}
  {et~al.}}{2010}]{ilbert+10}
{Ilbert} O.,  {et~al.} 2010, \apj, 709, 644

\bibitem[\protect\citeauthoryear{{Kannappan} \& {Gawiser}}{{Kannappan} \&
  {Gawiser}}{2007}]{kannappan_gawiser_07}
{Kannappan} S.~J.,  {Gawiser} E.,  2007, \apjl, 657, L5

\bibitem[\protect\citeauthoryear{{Korn}, {Maraston} \& {Thomas}}{{Korn}
  et~al.}{2005}]{korn+05}
{Korn} A.~J.,  {Maraston} C.,    {Thomas} D.,  2005, \aap, 438, 685

\bibitem[\protect\citeauthoryear{{Kriek}, {Labb{\'e}}, {Conroy}, {Whitaker},
  {van Dokkum}, {Brammer}, {Franx}, {Illingworth}, {Marchesini}, {Muzzin},
  {Quadri} \& {Rudnick}}{{Kriek} et~al.}{2010}]{kriek+10}
{Kriek} M.,  {Labb{\'e}} I.,  {Conroy} C.,  {Whitaker} K.~E.,  {van Dokkum}
  P.~G.,  {Brammer} G.~B.,  {Franx} M.,  {Illingworth} G.~D.,  {Marchesini} D.,
   {Muzzin} A.,  {Quadri} R.~F.,    {Rudnick} G.,  2010, \apjl, 722, L64

\bibitem[\protect\citeauthoryear{{Kurucz}}{{Kurucz}}{1992}]{kurucz_92}
{Kurucz} R.~L.,  1992, in {B.~Barbuy \& A.~Renzini} ed., The Stellar
  Populations of Galaxies Vol.~149 of IAU Symposium, {Model Atmospheres for
  Population Synthesis}.
p.~225

\bibitem[\protect\citeauthoryear{{Lan{\c c}on} \& {Mouhcine}}{{Lan{\c c}on} \&
  {Mouhcine}}{2002}]{lancon_mouhcine_02}
{Lan{\c c}on} A.,  {Mouhcine} M.,  2002, \aap, 393, 167

\bibitem[\protect\citeauthoryear{{Lan{\c c}on}, {Mouhcine}, {Fioc} \&
  {Silva}}{{Lan{\c c}on} et~al.}{1999}]{lancon+99}
{Lan{\c c}on} A.,  {Mouhcine} M.,  {Fioc} M.,    {Silva} D.,  1999, \aap, 344,
  L21

\bibitem[\protect\citeauthoryear{{Lawrence} {et~al.}}{{Lawrence}
    {et~al.}}{2007}]{lawrence+07} {Lawrence} A., {et~al.} 2007,
  \mnras, 379, 1599

\bibitem[\protect\citeauthoryear{{Lyubenova}, {Kuntschner}, {Rejkuba}, {Silva},
  {Kissler-Patig}, {Tacconi-Garman} \& {Larsen}}{{Lyubenova}
  et~al.}{2010}]{lyubenova+10}
{Lyubenova} M.,  {Kuntschner} H.,  {Rejkuba} M.,  {Silva} D.~R.,
  {Kissler-Patig} M.,  {Tacconi-Garman} L.~E.,    {Larsen} S.~S.,  2010, \aap,
  510, A19

\bibitem[\protect\citeauthoryear{{Lyubenova}, {Kuntschner}, {Rejkuba},
    {Silva}, {Kissler-Patig} \& {Tacconi-Garman}}{{Lyubenova}
    et~al.}{2012}]{lyubenova+12} {Lyubenova} M., {Kuntschner} H.,
  {Rejkuba} M., {Silva} D.~R., {Kissler-Patig} M., {Tacconi-Garman}
  L.~E., 2012, \aap, 543, A75

\bibitem[\protect\citeauthoryear{{MacArthur}, {McDonald}, {Courteau} \&
  {Jes{\'u}s Gonz{\'a}lez}}{{MacArthur} et~al.}{2010}]{macarthur+10}
{MacArthur} L.~A.,  {McDonald} M.,  {Courteau} S.,    {Jes{\'u}s Gonz{\'a}lez}
  J.,  2010, \apj, 718, 768

\bibitem[\protect\citeauthoryear{{Maraston}}{{Maraston}}{2005}]{maraston05}
{Maraston} C.,  2005, \mnras, 362, 799

\bibitem[\protect\citeauthoryear{{Maraston}, {Daddi}, {Renzini}, {Cimatti},
  {Dickinson}, {Papovich}, {Pasquali} \& {Pirzkal}}{{Maraston}
  et~al.}{2006}]{maraston+06}
{Maraston} C.,  {Daddi} E.,  {Renzini} A.,  {Cimatti} A.,  {Dickinson} M.,
  {Papovich} C.,  {Pasquali} A.,    {Pirzkal} N.,  2006, \apj, 652, 85

\bibitem[\protect\citeauthoryear{{Maraston} \& {Str{\"o}mb{\"a}ck}}{{Maraston}
  \& {Str{\"o}mb{\"a}ck}}{2011}]{maraston_stroembaeck_11}
{Maraston} C.,  {Str{\"o}mb{\"a}ck} G.,  2011, \mnras, 418, 2785

\bibitem[\protect\citeauthoryear{{Marigo} \& {Girardi}}{{Marigo} \&
  {Girardi}}{2007}]{marigo_girardi07}
{Marigo} P.,  {Girardi} L.,  2007, \aap, 469, 239

\bibitem[\protect\citeauthoryear{{Marigo}, {Girardi}, {Bressan}, {Groenewegen},
  {Silva} \& {Granato}}{{Marigo} et~al.}{2008}]{marigo+08}
{Marigo} P.,  {Girardi} L.,  {Bressan} A.,  {Groenewegen} M.~A.~T.,  {Silva}
  L.,    {Granato} G.~L.,  2008, \aap, 482, 883

\bibitem[\protect\citeauthoryear{{Meidt} {et~al.}}{{Meidt}
  {et~al.}}{2012}]{meidt+12}
{Meidt} S.~E.,  {et~al.} 2012, ArXiv e-prints

\bibitem[\protect\citeauthoryear{{Melbourne}, {Williams}, {Dalcanton},
  {Rosenfield}, {Girardi}, {Marigo}, {Weisz}, {Dolphin}, {Boyer}, {Olsen},
  {Skillman} \& {Seth}}{{Melbourne} et~al.}{2012}]{melbourne+12}
{Melbourne} J.,  {Williams} B.~F.,  {Dalcanton} J.~J.,  {Rosenfield} P.,
  {Girardi} L.,  {Marigo} P.,  {Weisz} D.,  {Dolphin} A.,  {Boyer} M.~L.,
  {Olsen} K.,  {Skillman} E.,    {Seth} A.~C.,  2012, \apj, 748, 47

\bibitem[\protect\citeauthoryear{{Moorwood} {et~al.}}{{Moorwood}
  {et~al.}}{1998}]{ISAAC}
{Moorwood} A.,  {et~al.} 1998, The Messenger, 94, 7

\bibitem[\protect\citeauthoryear{{O'Donnell}}{{O'Donnell}}{1994}]{odonnell_94}
{O'Donnell} J.~E.,  1994, \apj, 422, 158

\bibitem[\protect\citeauthoryear{{Pracy}, {Owers}, {Couch}, {Kuntschner},
  {Bekki}, {Briggs}, {Lah} \& {Zwaan}}{{Pracy} et~al.}{2012}]{pracy+12}
{Pracy} M.~B.,  {Owers} M.~S.,  {Couch} W.~J.,  {Kuntschner} H.,  {Bekki} K.,
  {Briggs} F.,  {Lah} P.,    {Zwaan} M.,  2012, \mnras, 420, 2232

\bibitem[\protect\citeauthoryear{{Quintero}, {Hogg}, {Blanton}, {Schlegel},
  {Eisenstein}, {Gunn}, {Brinkmann}, {Fukugita}, {Glazebrook} \&
  {Goto}}{{Quintero} et~al.}{2004}]{quintero+04}
{Quintero} A.~D.,  {Hogg} D.~W.,  {Blanton} M.~R.,  {Schlegel} D.~J.,
  {Eisenstein} D.~J.,  {Gunn} J.~E.,  {Brinkmann} J.,  {Fukugita} M.,
  {Glazebrook} K.,    {Goto} T.,  2004, \apj, 602, 190

\bibitem[\protect\citeauthoryear{{Riffel}, {Pastoriza},
  {Rodr{\'{\i}}guez-Ardila} \& {Maraston}}{{Riffel} et~al.}{2007}]{riffel+07}
{Riffel} R.,  {Pastoriza} M.~G.,  {Rodr{\'{\i}}guez-Ardila} A.,    {Maraston}
  C.,  2007, \apjl, 659, L103

\bibitem[\protect\citeauthoryear{{Schlegel}, {Finkbeiner} \&
  {Davis}}{{Schlegel} et~al.}{1998}]{schlegel_dust}
{Schlegel} D.~J.,  {Finkbeiner} D.~P.,    {Davis} M.,  1998, \apj, 500, 525

\bibitem[\protect\citeauthoryear{{Srinivasan} {et~al.}}{{Srinivasan}
  {et~al.}}{2009}]{srinivasan+09}
{Srinivasan} S.,  {et~al.} 2009, \aj, 137, 4810

\bibitem[\protect\citeauthoryear{{Strauss} {et~al.}}{{Strauss}
  {et~al.}}{2002}]{strauss_etal02}
{Strauss} M.~A.,  {et~al.} 2002, \aj, 124, 1810

\bibitem[\protect\citeauthoryear{{Thomas}, {Maraston} \& {Bender}}{{Thomas}
  et~al.}{2003}]{thomas+03}
{Thomas} D.,  {Maraston} C.,    {Bender} R.,  2003, \mnras, 339, 897

\bibitem[\protect\citeauthoryear{{Thomas}, {Maraston}, {Bender} \& {Mendes de
  Oliveira}}{{Thomas} et~al.}{2005}]{thomas+05}
{Thomas} D.,  {Maraston} C.,  {Bender} R.,    {Mendes de Oliveira} C.,  2005,
  \apj, 621, 673

\bibitem[\protect\citeauthoryear{{Thomas}, {Maraston} \& {Korn}}{{Thomas}
  et~al.}{2004}]{thomas+04}
{Thomas} D.,  {Maraston} C.,    {Korn} A.,  2004, \mnras, 351, L19

\bibitem[\protect\citeauthoryear{{Wright} {et~al.}}{{Wright}
  {et~al.}}{2010}]{wright+10}
{Wright} E.~L.,  {et~al.} 2010, \aj, 140, 1868

\bibitem[\protect\citeauthoryear{{York} {et~al.}}{{York}
  {et~al.}}{2000}]{SDSS}
{York} D.~G.,  {et~al.} 2000, \aj, 120, 1579

\bibitem[\protect\citeauthoryear{{Zibetti}, {Charlot} \& {Rix}}{{Zibetti}
  et~al.}{2009}]{ZCR09}
{Zibetti} S.,  {Charlot} S.,    {Rix} H.,  2009, \mnras, 400, 1181

\end{thebibliography}

\label{lastpage}
\clearpage
%
\end{document}